\documentclass[pdftex,twocolumn,epjc3]{svjour3}          % twocolumn
\RequirePackage[T1]{fontenc}
\smartqed  % flush right qed marks, e.g. at end of proof
\RequirePackage{graphicx}
\RequirePackage{mathptmx}      % use Times fonts if available on your TeX system
\RequirePackage{flushend}
\RequirePackage[numbers,sort&compress]{natbib}
\RequirePackage[colorlinks,citecolor=blue,urlcolor=blue,linkcolor=blue]{hyperref}
\RequirePackage{amsmath,amssymb}
\RequirePackage{multirow}
\RequirePackage{xspace}
\RequirePackage{xcolor}
\RequirePackage{cancel}
\RequirePackage[caption=false]{subfig}
\RequirePackage{subfig}
\RequirePackage{blindtext, subfig}
\RequirePackage{amsmath}
\usepackage{ulem}

\RequirePackage{comment}
\RequirePackage{hyperref}
\RequirePackage{slashed}
\RequirePackage[caption=false]{subfig}
\RequirePackage{tabu}
\usepackage{hyperref}
\usepackage{multicol}
\allowdisplaybreaks

\journalname{}

\newcommand{\ba}{  \begin{array}}
\newcommand{\ea}{\end{array}}

\newcommand{\bd}{  \begin{displaymath}}
\newcommand{\ed}{\end{displaymath}}

\newcommand{\bsube}{  \begin{subequation}}
\newcommand{\esube}{\end{subequation}}
\newcommand{ \bea}{  \begin{eqnarray}}
\newcommand{ \eea}{\end{eqnarray}}
\newcommand{\bal}{  \begin{align}}
\newcommand{\ealign}{\end{align}}
\newcommand{\eal}{\end{align}}
\newcommand{ \bean}{  \begin{enumerate}}
\newcommand{ \eean}{\end{enumerate}}

\begin{document}

\title{Measuring $CP$ nature of $h\tau{\bar\tau}$ coupling at $e^-p$ collider}
%\subtitle{Do you have a subtitle?\\ If so, write it here}

\author{Sukanta Dutta\thanksref{e1, addr1} 
             \and Ashok Goyal\thanksref{e2, addr2}
             \and Mukesh Kumar\thanksref{e3, addr3}
             \and Abhaya Kumar Swain\thanksref{e4, addr1,addr2,addr3}
}

%\thankstext[$\star$]{t1}{Thanks to the title}
\thankstext{e1}{e-mail: Sukanta.Dutta@gmail.com}
\thankstext{e2}{e-mail: agoyal45@yahoo.com}
\thankstext{e3}{e-mail: mukesh.kumar@cern.ch}
\thankstext{e4}{e-mail: abhaya.kumar.swain@cern.ch}

\institute{SGTB Khalsa College, University of Delhi (DU), Delhi, India.\label{addr1}
\and
Department of Physics \& Astrophysics, University of Delhi, Delhi, India.\label{addr2}
\and
School of Physics and Institute for Collider Particle Physics, University of the Witwatersrand, Johannesburg, Wits 2050, South Africa.\label{addr3}
}

\date{}
% The correct dates will be entered by the editor

\maketitle

\begin{abstract}
The proposed future $e^- p$ collider provides sufficient energies to produce the Standard Model Higgs Boson ($h$) through $W^\pm$ and $Z$-Boson fusion in charged and neutral current modes, respectively and to measure its properties. We take this opportunity to investigate the prospect of measuring the $CP$  properties of $h$ through $h \to \tau^+ \tau^-$, where $\tau^-\,\left(\tau^+\right)$ decays to a charged pion $\pi^-\,\left(\pi^+\right)$ and a neutral pion $\pi^0$ in association with  neutrino (anti-neutrino). An interesting $CP$ sensitive angular observable $\alpha_{\rm CP}$ between the two $\tau$-leptons decay plane in the $\pi^+\pi^-$ centre of mass frame is proposed and investigated in this work. For fixed electron energy of 150~GeV along with 7 (50)~TeV of proton energy, the $CP$ phase can be measured approximately to $25^\circ$ ($14^\circ$) at integrated luminosity of 1~ab$^{-1}$ for $-$80\% polarised electron at 95\% confidence level.    
\end{abstract}

\section{Introduction}
\label{intro}
After the Higgs Boson ($h$) discovery at the Large Hadron Collider (LHC)~\cite{ATLAS:2013xga, ATLAS:2012yve, CMS:2012qbp, ATLAS:2013dos}, measurement of its properties and their possible deviation from the Standard Model (SM) predictions are important to explore physics beyond the SM (BSM)~\cite{CMS:2020dkv}. In this context the proposed Large Hadron electron Collider (LHeC)~\cite{AbelleiraFernandez:2012cc,Bruening:2013bga,LHeC:2020van, FCC:2018byv} with centre of mass (CM) energy $\sqrt{s} \approx 1.3$~TeV with possible enhancement to 3.5~TeV at the proposed Future Circular Electron Hadron Collider  (FCC-eh) programme at CERN may act as potential Higgs factories offering enormous scope to study the Higgs Boson properties~\cite{Biswal:2012mp,Kumar:2015kca,Coleppa:2017rgb}. Exploring $CP$ nature of the $h$ coupling to the SM particles as occurs in several extensions of the SM through the interaction of multiple Higgs-sector, in super-symmetric theories and so on, has become important since $CP$ violation in the Higgs Boson sector would impact baryogensis in the early Universe  \cite{Ge:2020mcl,Basler:2017uxn,Bernlochner:2018opw,Shu:2013uua,Chiang:2016vgf}. 

\par The ATLAS and CMS collaborations have probed the $CP$ nature of the $h$ coupling and have excluded the pure $CP$ odd nature at 99\% confidence level (C.L.)   ~\cite{CMS:2012vby, CMS:2013fjq, CMS:2014nkk, CMS:2016tad}, and this leaves the possibility of $h$ either being an admixture of $CP$-even and $CP$-odd states or a pure $CP$-even state. Current bounds on the mixing angle are weak, and large mixing is not ruled out. This property has been analysed in a clean di-lepton pair production through $h \to Z Z^* \to 4 \ell $~\cite{Chen:2014gka, Bishara:2013vya, Korchin:2013ifa, Chen:2014ona}. However, this process has low sensitivity to determine the $CP$ violating phase because of the dominant $CP$ even $hZZ$ coupling. The $CP$-odd scalar coupling to $Z/W^\pm$ vector Bosons can arise only from the dimension six SM gauge invariant operators and are likely to be subdominant as compared to the $CP$-even SM couplings. Since the Yukawa coupling of $h$ to the third generation fermions is larger, it is natural to expect that studying $CP$ properties with them might play an important role. In Ref.   ~\cite{Coleppa:2017rgb}, authors studied the $h$ coupling with top-quark in the LHeC environment. 

It is also important to mention that the choice of $CP$-odd observable are often sensitive to the  production mechanism of Higgs Boson. The dominant gluon fusion channel at the LHC suffers from alarmingly large SM background   ~\cite{Harnik:2013aja, Dolan:2014upa} which, in turn, reduce the signal rate. In contrast, $h$ production through vector-Boson fusion (VBF) has clear advantage over the gluon-fusion mode. The VBF leads to distinctive topology resulting in an enhancement of the signal to background ratio~\cite{Rainwater:1998kj}.

In the context of measuring $CP$ characteristics of Higgs Boson through the $\tau$-lepton Yukawa coupling, several angular observables has been proposed in the literature at the collider experiments in Refs.   ~\cite{DellAquila:1985jin, DellAquila:1988bko, Bernreuther:1993df, Bernreuther:1993hq, Soni:1993jc, Skjold:1994qn, Grzadkowski:1995rx, Grzadkowski:1999ye, Hagiwara:2000tk, Han:2000mi, Plehn:2001nj, Choi:2002jk, Bower:2002zx, Desch:2003mw, Asakawa:2003dh, Desch:2003rw, Godbole:2004xe, Rouge:2005iy, Biswal:2005fh, Ellis:2005ika, Godbole:2007cn, BhupalDev:2007ftb, Berge:2008wi, Berge:2008dr, DeRujula:2010ys, Christensen:2010pf, Berge:2011ij, Godbole:2011hw, Harnik:2013aja, Berge:2013jra, Chen:2014ona, Dolan:2014upa, Hayreter:2015cia, Han:2016bvf, Bhardwaj:2016lcu, Swain:2020sil}. These observables are defined in the $\tau$-lepton pair centre of mass frame. However, reconstructing the CM frame of $h$ in $\tau^+\tau^-$-channel is extremely challenging. Firstly, because of the unknown CM energy of collision between the incoming partons in hadron colliders and secondly, the presence of the neutrinos which escape the detector leaving the transverse momentum imbalance to infer their presence. Since there are multiple neutrinos present in the final state, it is very difficult to reconstruct the $\tau^\pm$ momenta. However, it is well known that the $\tau$-lepton has finite decay length which results in a measurable impact parameter. This additional measurement not only improve the $\tau^\pm$ momentum reconstruction   ~\cite{Hagiwara:2016zqz, Bhardwaj:2016lcu} but also leads to construct the $CP$ sensitive angular observables   ~\cite{Berge:2008dr, Berge:2014sra} without even requiring the reconstruction of CM frame of $h$. One such angular observable has been explored to measure the $CP$ phase of the $\tau$-lepton Yukawa coupling at the CMS collaboration   ~\cite{CMS:2020rpr}. This measurement constraint the $CP$ phase to $4^\circ \pm 17^\circ$ ($\pm 36^\circ$)  at 68\% (95\%) C.L. with CM energy of 13~TeV at integrated luminosity, $\cal{L} =$  137~fb$^{-1}$. 

\begin{figure}[t]%[!htbp]
  \centering
  \includegraphics[width=0.45\textwidth]{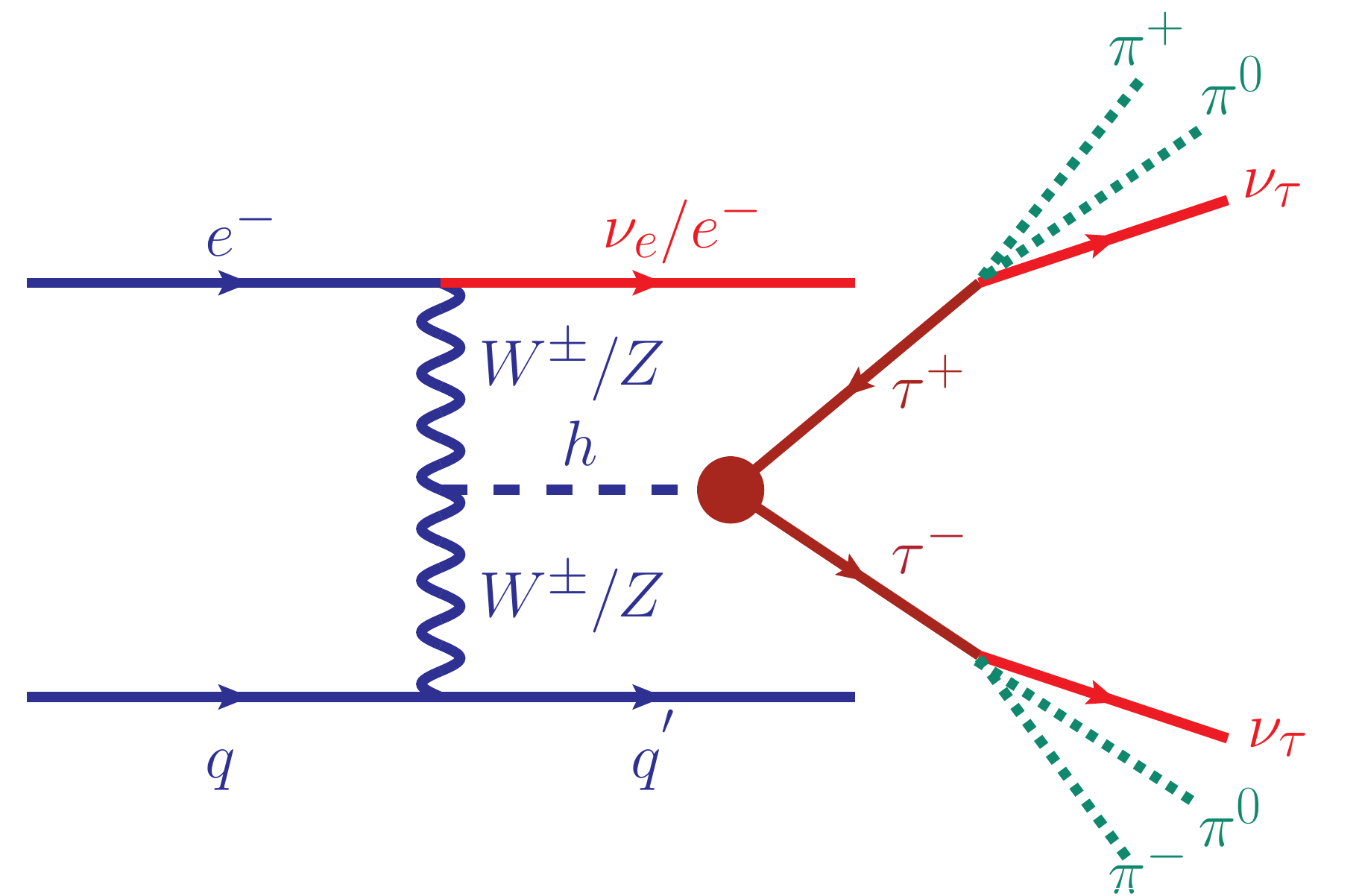}\label{fig:figNc}
  \caption{  \small Leading order Feynman diagram for $h$ production {\it via} the charged / neutral current process in $e^- p$ collider. The figure shows the process $e^- p \to \nu_e/e^- h \,j$, $h \to \tau^+ \tau^-$ with $\tau^\pm \to \pi^\pm \pi^0 \nu_\tau$, $q$ are the partons from proton and $q^\prime$ is the scattered jets.}
\label{fig:figW}
\end{figure}

In this article we focus on $h$ production at the proposed future $e^- p$ colliders, namely, the LHeC and FCC-eh and will probe the $CP$ nature of the $h\tau{\bar\tau}$ coupling in this environment. We will discuss the involved challenges in the analysis and probe one of the observable as mentioned above. The Higgs Boson is produced through the charged current (CC) $W^+W^-$ fusion and the neutral current (NC) $ZZ$ fusion giving rise to a forward jet and missing energy in the former and a forward jet with scattered electron in the latter as shown in Fig.~\ref{fig:figNc} with $h$ decaying to $\tau^+ \tau^-$ pair. Further we choose $\tau^\pm$ decay to $\rho^\pm$ with corresponding $\nu_\tau$ and $\rho^\pm$ decays to $\pi^\pm \pi^0$ with approximately 100\% probability.
We have picked up this channel among all other $\tau^\pm$ decay channels because of the larger branching ratio of 25.9$\%$ to demonstrate the prospect of measuring the $CP$ phase of the $h\tau{\bar\tau}$ coupling.
In this study we only modify the $h\tau{\bar\tau}$ vertex by assuming other involved couplings to be the SM one. 

In section~\ref{sec2:TheorysetUpAndSimulation} we discuss the Lagrangian which is used to parameterize the $CP$ phase dependent vertex of the $h\tau{\bar\tau}$. Then we describe the simulation that we follow to perform the analysis. Next in section~\ref{obsv} we elucidate the observable that is employed to constrain the $CP$ admixture of the $h\tau{\bar\tau}$ coupling. In the following section~\ref{Results}, we illustrate the results and section~\ref{summaryAndConclusion} is devoted for summary and discussions of our analysis.

\section{Formalism \& Simulation}
\label{sec2:TheorysetUpAndSimulation}

In this analysis we assume the measured Higgs Boson mass at $m_h = 125$~GeV to be a mixture of $CP$ even and odd scalar and the interaction between $h$ and $\tau^\pm$ is given by
\begin{equation}
{\cal L}_{h\tau\bar\tau} = -\left(\sqrt{2}\,G_F\right)^{1/2}m_\tau\,\,\bar\tau \left(\tilde{a} + i\tilde{b} \gamma_5 \right)\tau \,h; \,\, (\tilde a > 0) \label{LagDecay1}
\end{equation}
where  $G_F$ is the Fermi constant, $m_{\tau}$ is the mass of $\tau$-leptons and the parameters $\tilde{a}$ ($\tilde{b}$) are  dimensionless couplings for the $CP$ even (odd) part of the Lagrangian. In, an alternative parametrization, the Lagrangian in equation~\eqref{LagDecay1} can be expressed as 
\begin{equation}
{\cal L}_{h\tau\bar\tau} = -C^\tau_{\rm eff.}\,\bar\tau \left(\cos\phi_\tau + i\sin\phi_\tau \gamma_5 \right)\tau \,h, \label{LagDecay2}
\end{equation}
where the $C^\tau_{\rm eff.}\equiv\left(\sqrt{2}\,G_F\right)^{1/2}m_\tau\,\sqrt{\tilde a^2+\tilde b^2} $ is the effective coupling and $\phi_\tau\equiv-{\pi}/{2}\le\tan^{-1}\left({\tilde b}/{\tilde a}\right)\le\pi/2$ %$\phi_\tau\equiv-\frac{\pi}{2}\le\tan^{-1}\left[ \frac{\tilde b}{\tilde a}\right]\le \frac{\pi}{2}$ 
is the mixing angle of the scalar and pseudo-scalar component of the couplings with leptons. The choices  $(\tilde a =1,\, \tilde b=0)$, $(\tilde a =1/\sqrt{2},\, \tilde b=1/\sqrt{2})$   and $(\tilde a =0,\, \tilde b=1)$  correspond to SM pure CP-even ($\phi_\tau=0^\circ$), a 50\% mixed ($\phi_\tau=45^\circ$) and pure CP-odd pseudo-scalar ($\phi_\tau=90^\circ$) states  for both the parametrization respectively. Also, they satisfy $\sqrt{\tilde a^2 +\tilde b^2}$ = 1 implying that the effective strength of the Yukawa couplings $C^\tau_{\rm eff.}$ for  the mixed or pure states are identical to that of SM modulo the dependence on mixing angle. It is to be noted that the mixing angle $\phi_\tau$ is specific for the $\tau$ lepton and is not universal with respect to ({\it w.r.t.}) other SM fermions. In our analysis, we have assumed all other couplings of the fermions and gauge Bosons with the scalar $h$ to be identical to those of SM.

As mentioned in the introduction, we consider the following $\tau^\pm $  decay mode in our analysis: $\tau^{-(+)}\to\nu_\tau\,(\bar\nu_\tau) +\rho^{-(+)}\to \nu_\tau\,(\bar\nu_\tau)+\pi^{-(+)}+\pi^0$. The effective Lagrangian  describing the $\rho^\pm$  decay mode is given as 
%\begin{eqnarray}
\begin{align}
&{\cal L}_{\pi^\pm,\,\pi^0} = \sqrt{2}\, G_F\, f_2\, \,\,\bar \tau \gamma^\mu \,P_L\,\nu_\tau\,\left(\pi^-\,\partial_\mu\pi^--\pi^-\partial_\mu\pi^0\right) +{\rm h.c.}\notag\\ 
&{\rm with}\,\, f_2=\sqrt{2}\,\cos\theta_C\, F_\rho\left(Q^2\right).\label{taudecayvertex}
\end{align}
%\end{eqnarray}
The detailed discussions on parametrization of the form factor $F_\rho\left(Q^2\right)$ are found in references~\cite{Hagiwara:2012vz,Jadach:1990mz,Kuhn:1990ad}.

\par We have implemented the Lagrangians given in equations ~\eqref{LagDecay1} and~\eqref{taudecayvertex} in \texttt{FeynRules}~\cite{Alloul:2013bka} to build the model file and simulated the parton level events  using \texttt{MadGraph5}~\cite{Alwall:2014hca} with \texttt{NN23LO1}~\cite{Ball:2012cx} parton distribution function. The factorisation and renormalisation scales for the simulation are taken to be the default \texttt{MadGraph5} dynamic scales. Both the CC and NC channels are simulated in the LHeC set up with electron (proton) beam energy to be 150 (7000)~GeV and FCC-eh set up with proton beam energy of 50~TeV keeping other parameters unchanged. The analysis is performed for the unpolarised and - 80\% polarised electron beams, respectively. In addition, we require transverse momenta of charged pions $p_{T_\pi^\pm} \ge 20$~GeV.
\begin{table}[t]
\centering
\resizebox{0.85\linewidth}{!}{
{\tabulinesep=3pt
\begin{tabu}{|c|c|c|}
	\hline
	Process  & {LHeC} : $\sigma$(fb) & {FCC-eh} : $\sigma$(fb) \\
	\hline
	\hline
	Signal: CC  & 0.56 (1.0) & 1.72 (3.11) \\
	\hline
	Signal: NC  & 0.1 (0.11)  & 0.37 (0.41)\\
	%Bkg: $p e^- \rightarrow e^- j~\tau^- \tau^+$ $\backslash h$        & $2.04 \times 10^{1}$ \\ 
	\hline
\end{tabu}}
}
\caption{  {\small SM cross sections for the CC and NC processes as shown in Fig.~\ref{fig:figW} for unpolarised ($-$80\% polarised) 150~GeV electron beam colliding with 7~TeV and 50~TeV  proton beams corresponding to LHeC and FCC-eh respectively.}}
\label{tab:xsec}
\end{table}
\begin{figure}[t]
	\includegraphics[width=0.40\textwidth,height=0.40\textwidth]{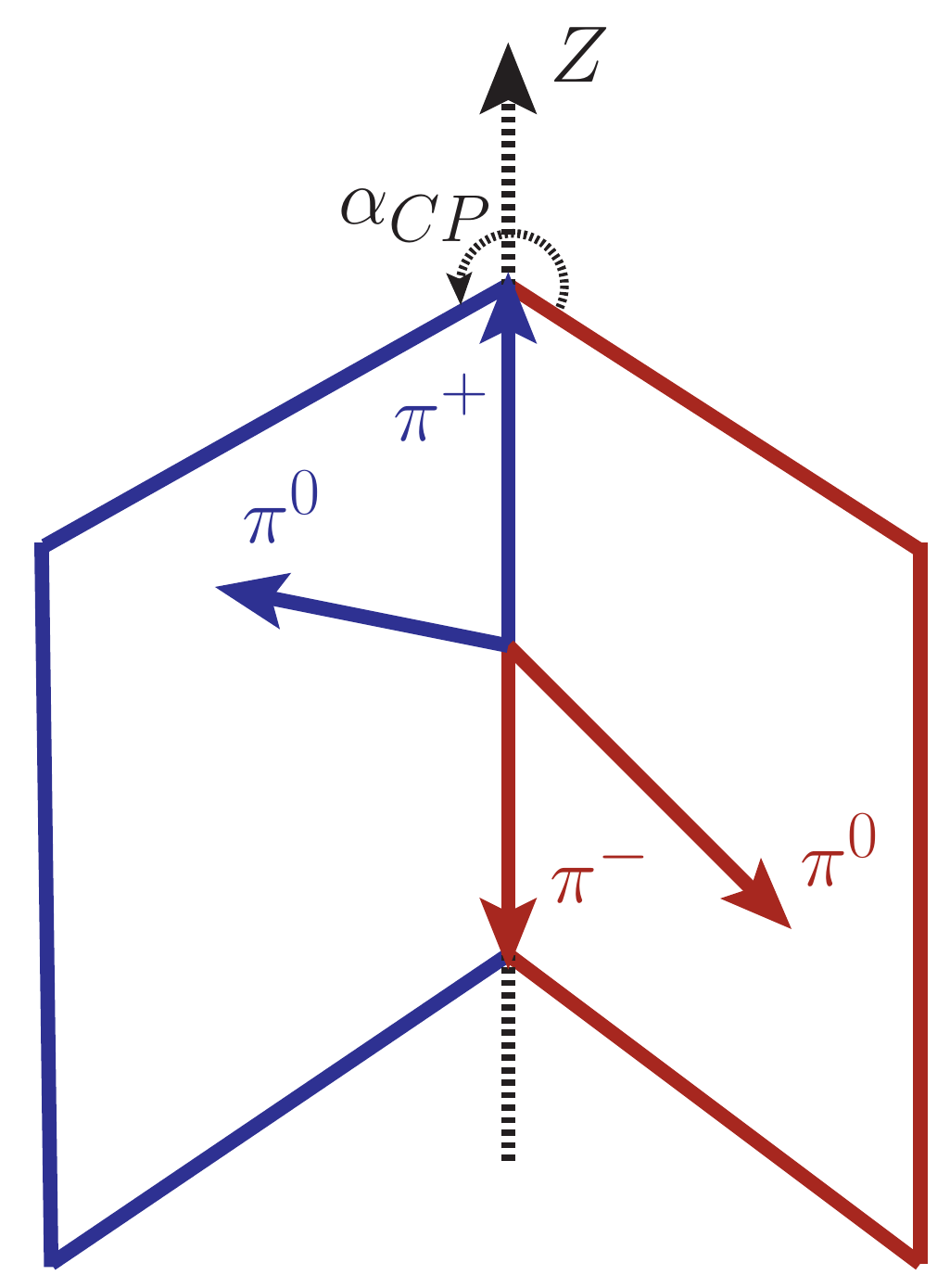}
	\caption{  {\small Representative diagram for the decay planes spanned by the charged and neutral pion produced from respective $\tau^\pm$-lepton in the $\pi^{+}\pi^{-}$ rest frame. The angle between the two decay planes is denoted as $\alpha_{CP}$ which is utilised here to explore the $CP$ admixture of the $h\tau{\bar\tau}$ coupling.}}
	\label{fig:decayPlane}
\end{figure}

In Table~\ref{tab:xsec}, the simulated CC and NC cross-sections are shown  for 150~GeV unpolarised and $-$80\% polarised electron beam colliding with 7~(50)~TeV proton beam corresponding to the proposed LHeC (FCC-eh) collider. Further, we discuss the $CP$ sensitivity of    $h\tau{\bar\tau}$ coupling in the next section.  

\par Before, concluding  this section, we briefly mention the potential backgrounds to CC and NC processes. The dominant SM tree level background for the CC process  comes from $W^+W^-$ fusion process $e^-p\to \nu_e\,j\,Z/Z^*/\gamma^*$, while backgrounds for NC process are induced by the $\gamma^\star Z\gamma$, $Z^\star Z \gamma $,  $\gamma^\star ZZ $   and $Z^\star\, ZZ$ vertices  either at the one loop level or  through the higher dimensional model independent effective operators. The $\tau^-\tau^+$ pair  emanating from the neutral on-shell/ off-shell gauge Bosons decay posses a different angular distribution when compared with those produced from the decay of the mixture of scalar and pseudo-scalar. This has a bearing on the angular distribution on the decay products $\tau^\pm$ in $\rho^\pm$ mode and can be suppressed with appropriate selection cuts. The contribution from the dominant on-shell $Z$-decay  can be vetoed by imposing a selection cut on the invariant mass of the $\tau^-\tau^+$ pairs. 
\begin{figure}[t]
	\includegraphics[width=0.5\textwidth,height=0.40\textwidth]{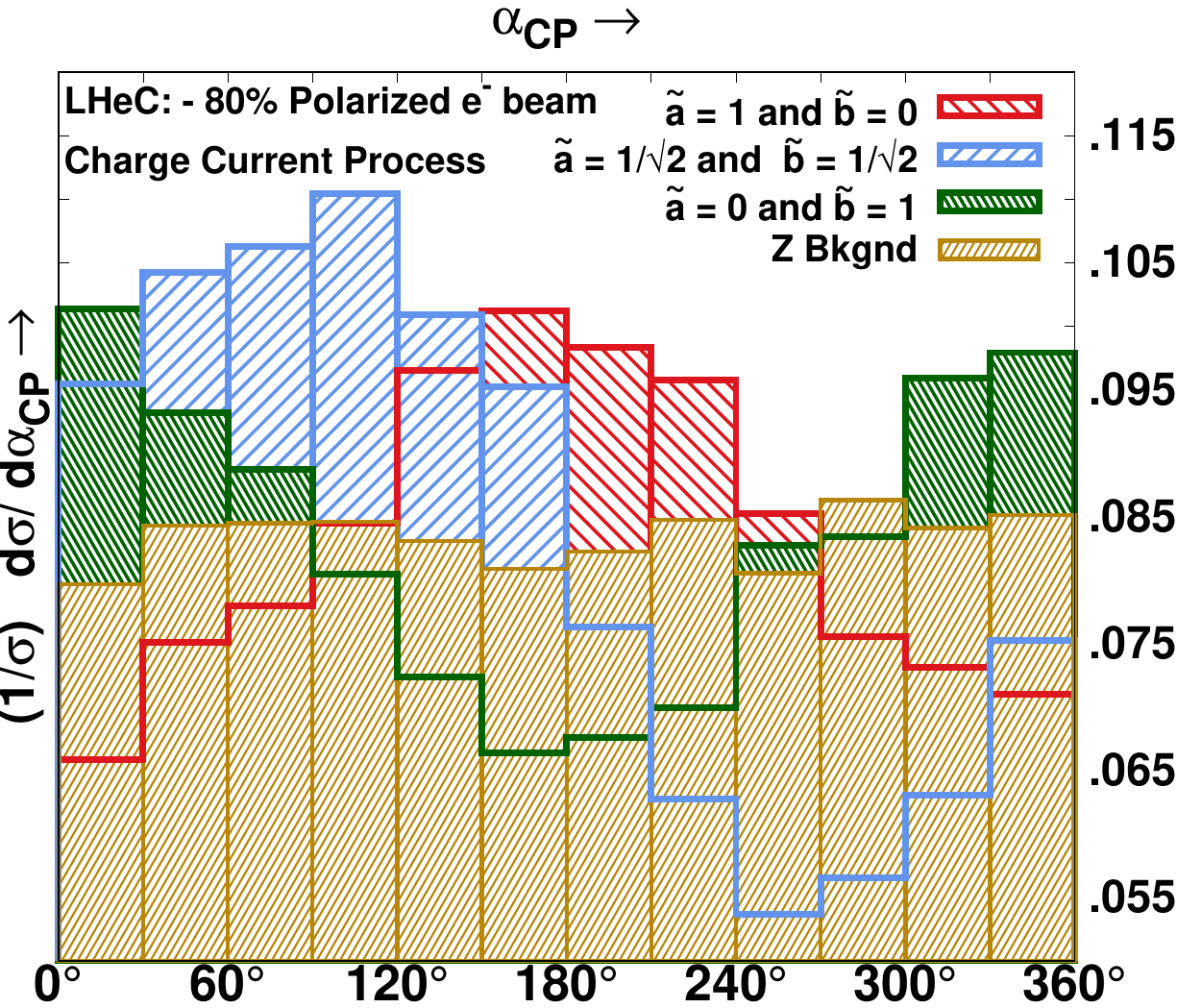}
	\caption{  {\small Normalized differential cross-sections   {\it  w.r.t.} $CP$ sensitive observable $\alpha_{CP}$  for the process  $e^-p\to (h\to \tau^+\tau^-)\, \nu_e\, {\rm jets}$; and $\tau^{-(+)}\to\nu_\tau\,(\bar\nu_\tau) +\rho^{-(+)}\to \nu_\tau\,(\bar\nu_\tau)+\pi^{-(+)}+\pi^0$ corresponding to SM pure $CP$-even (in red), 50\% mixed (in blue) and pure $CP$-odd  (in green) states respectively. The contribution from dominant $Z$ background channel is shown in golden yellow. These distributions are simulated for LHeC set up with $-$80\% polarised 150~GeV electron   and unpolarised 7~TeV proton beams.}}
	\label{fig:distLHeC}
\end{figure}
\begin{figure*}[t]
\centering
\subfloat[][]{\includegraphics[width=.5\textwidth]{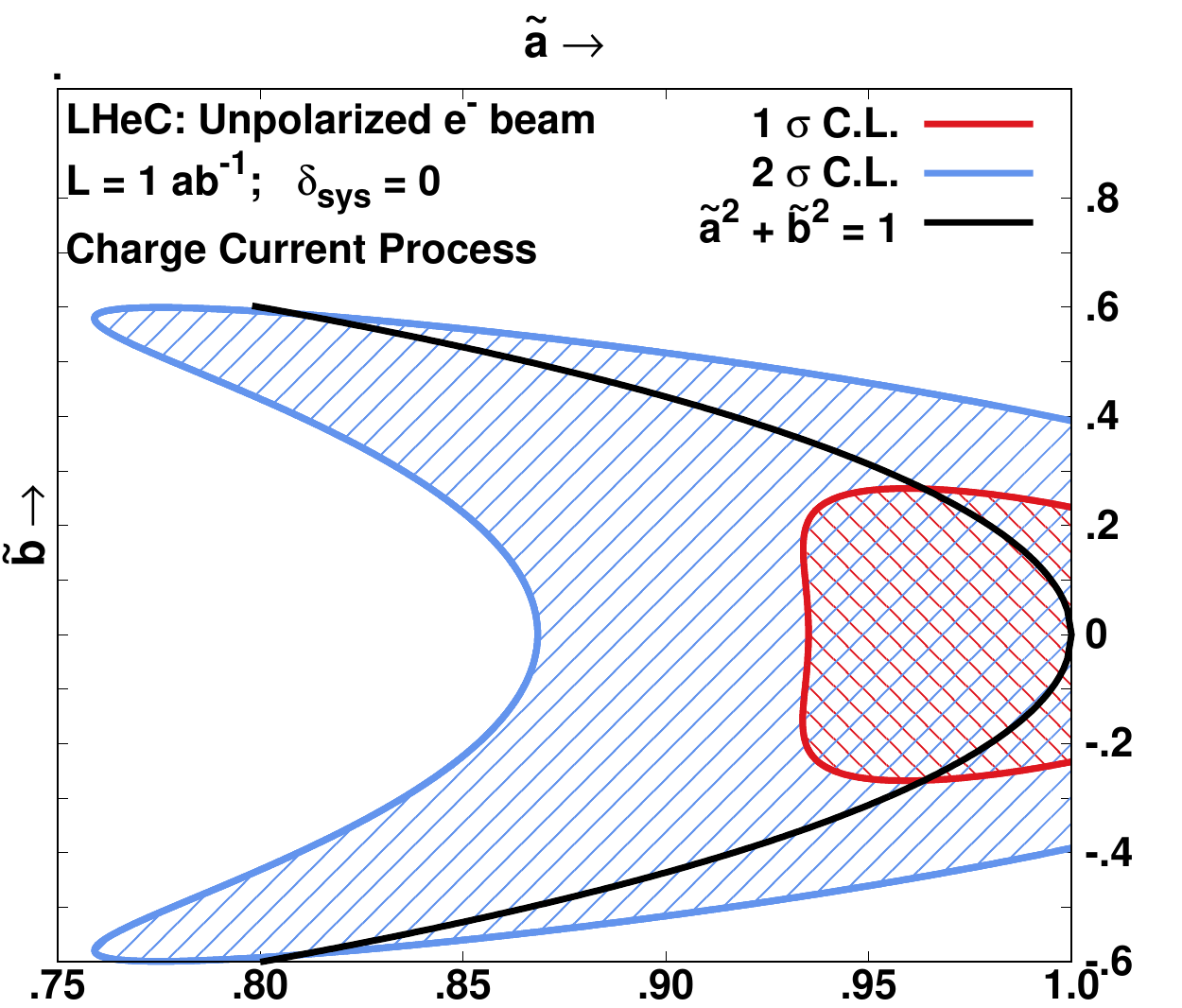}}
\subfloat[][]{\includegraphics[width=.5\textwidth]{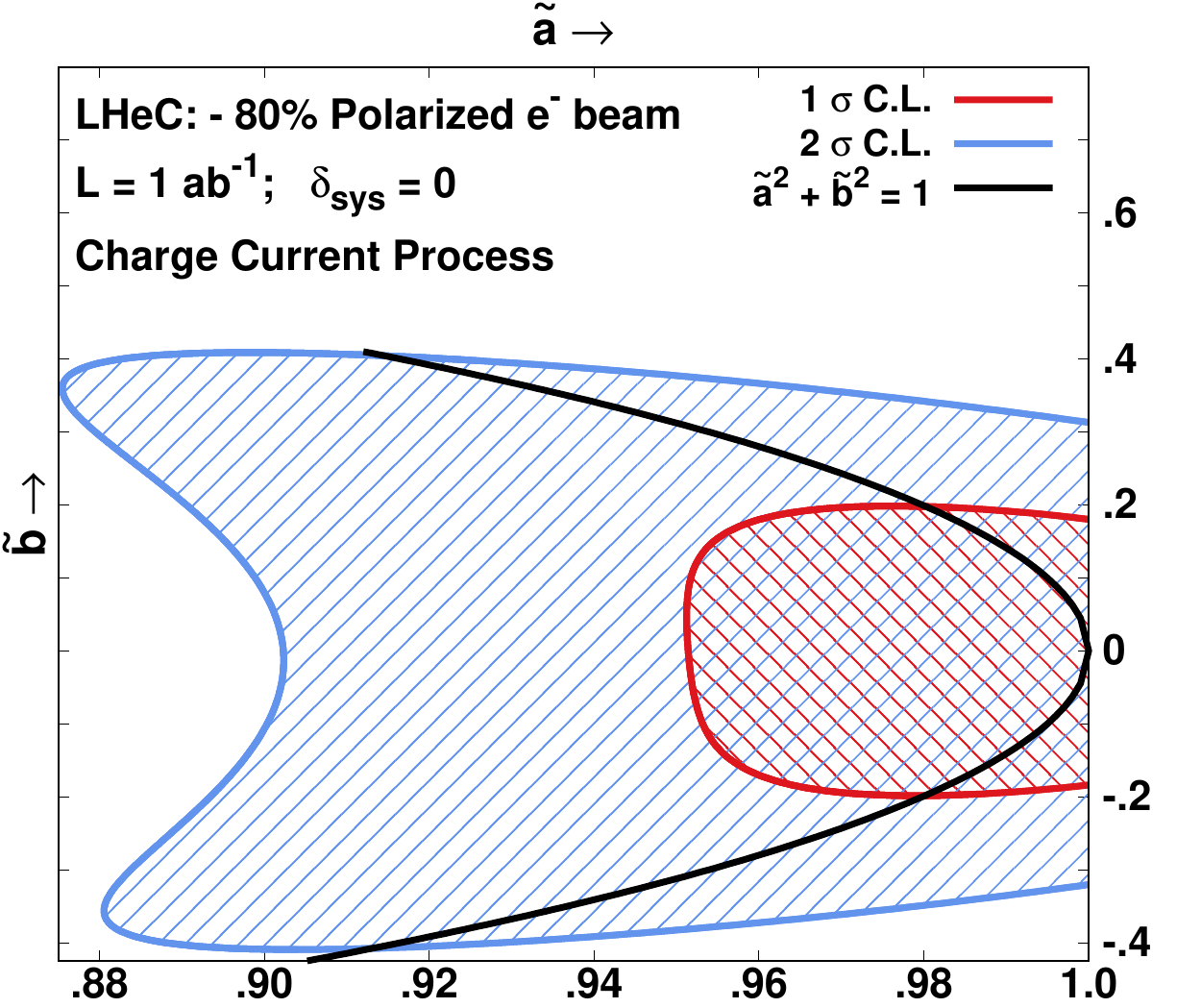}}
\caption{  \small 68\% (in red) and 95\% (in blue)   C.L. exclusion contours are drawn in  $\tilde a-\tilde b$ plane using differential bin-width $\Delta\alpha_{CP}$ = $30^\circ$. In each panel the  intersecting points of the black contour  with the one and two  sigma contours   depict the respective mixing angles $\phi_\tau$ with SM like Yukawa coupling strength. Shaded interiors of the respective contours are however remain insensitive to the collider.}
\label{fig:CC_Ee150}
\end{figure*}
\begin{figure*}[t]
	\centering
	\subfloat[][]{\includegraphics[width=.5\textwidth]{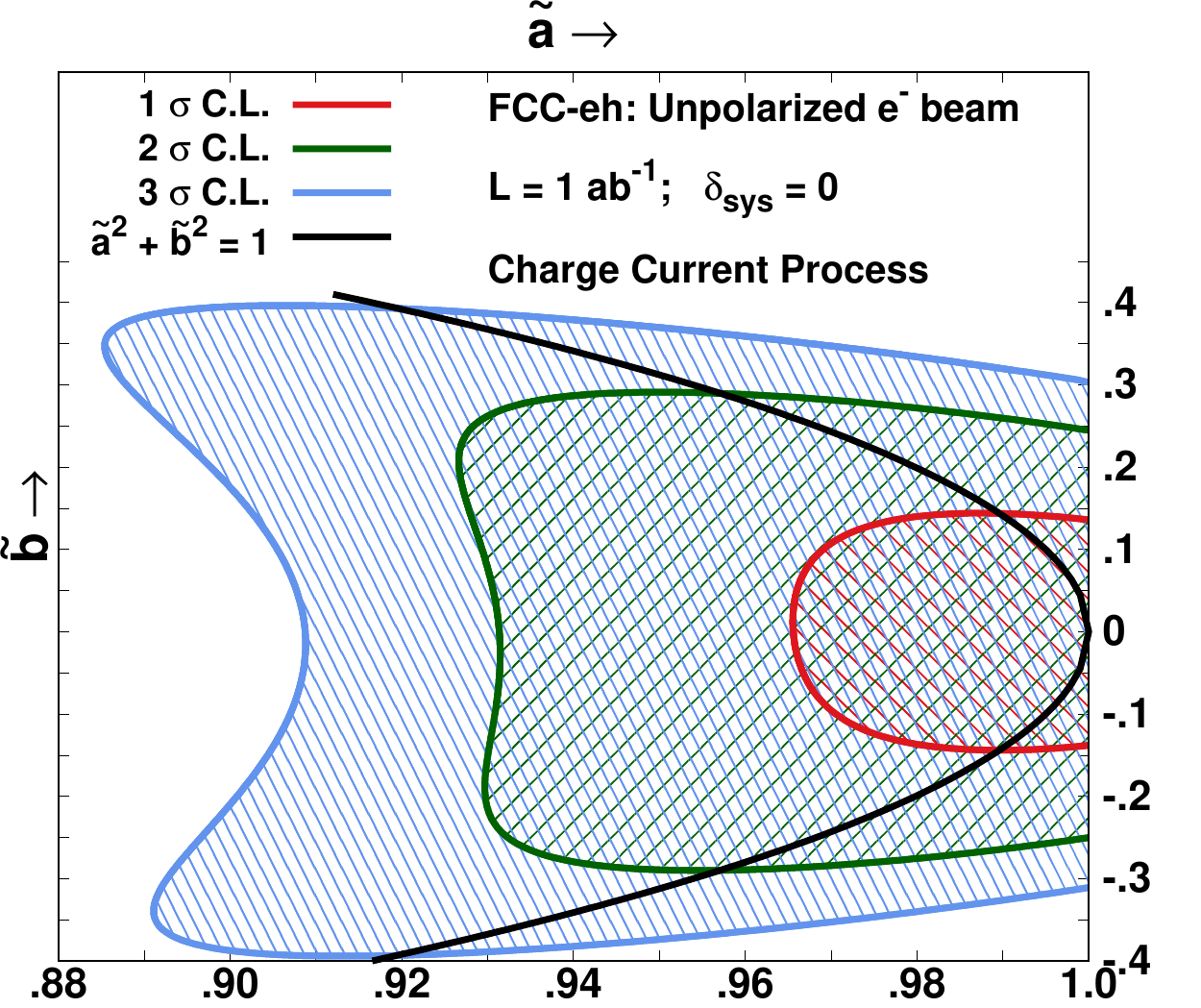}}
	\subfloat[][]{\includegraphics[width=.5\textwidth]{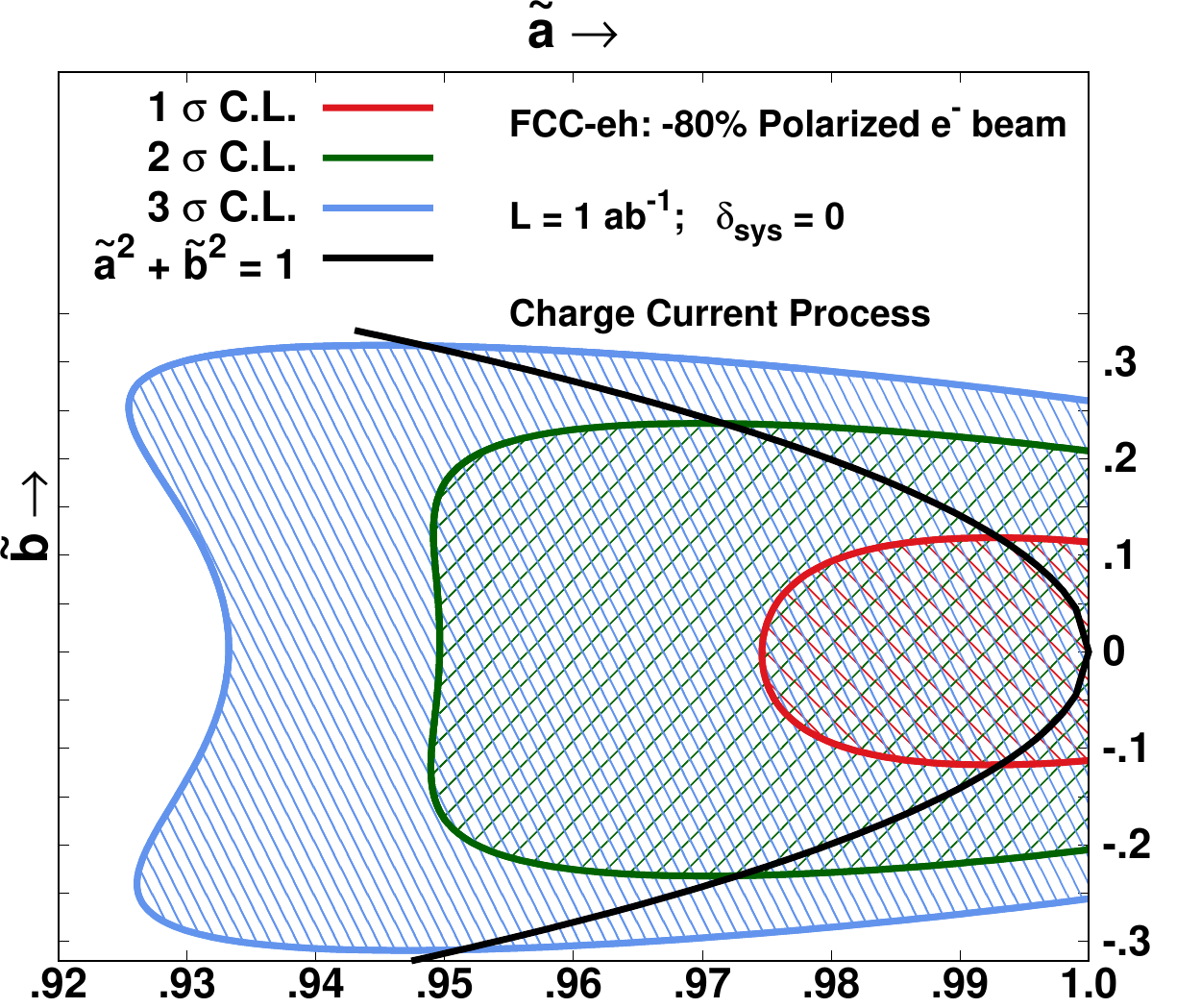}}\\
	\subfloat[][]{\includegraphics[width=.5\textwidth]{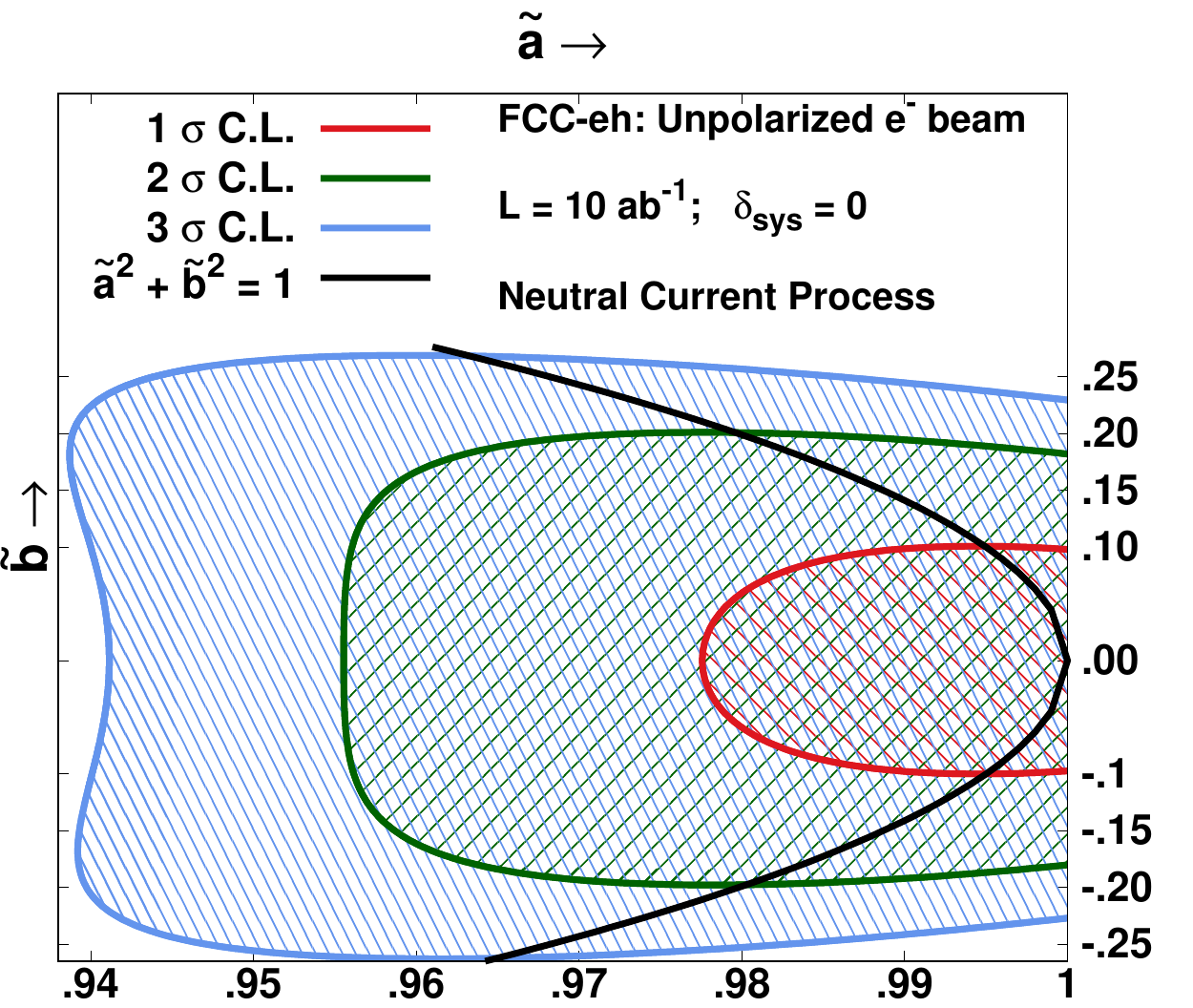}}
	\subfloat[][]{\includegraphics[width=.5\textwidth]{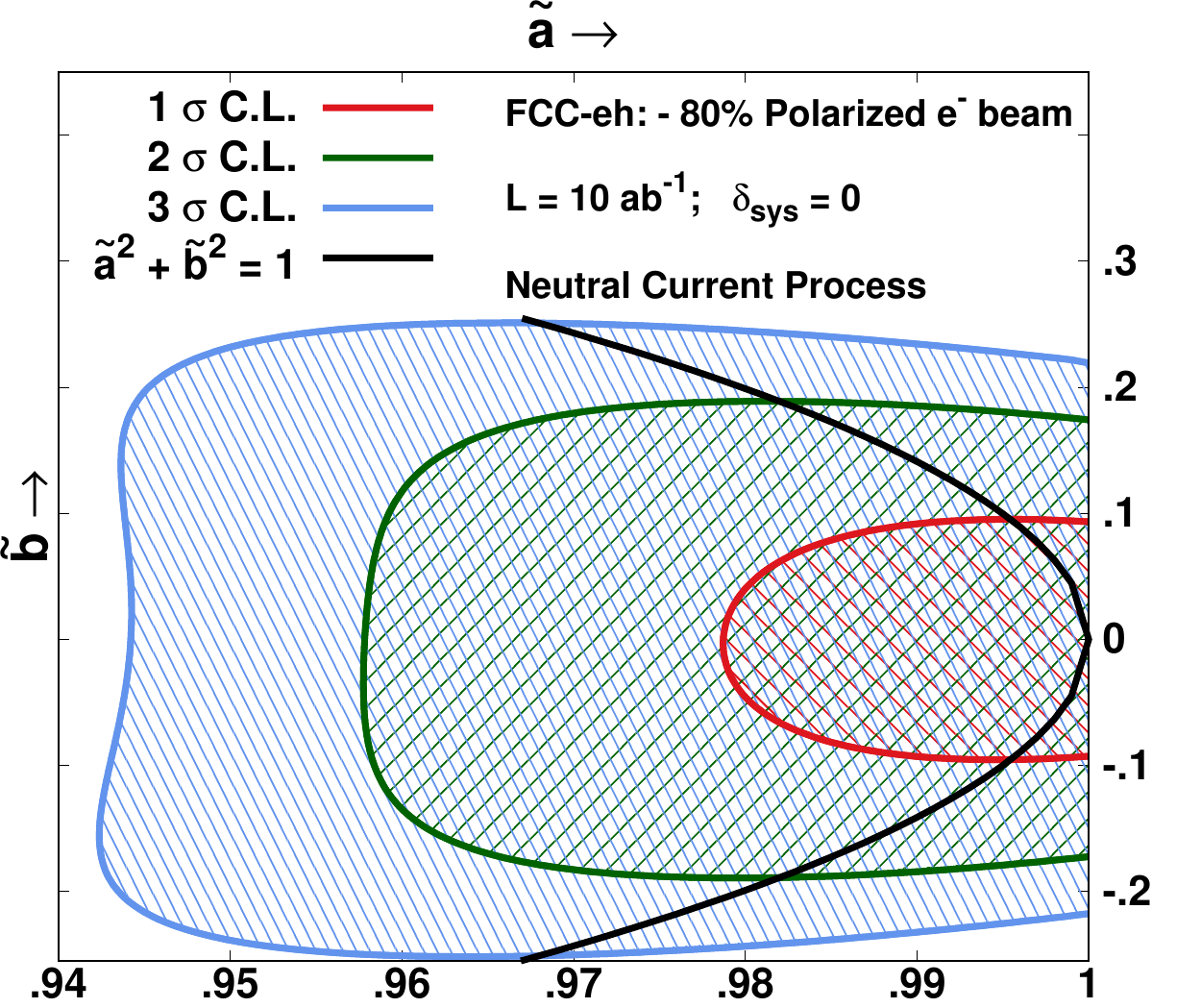}}
\caption{\small 68\% (in red), 95\% (in green) and 99.7\% (in blue)  C.L. exclusion contours are drawn in  $\tilde a-\tilde b$ plane using differential  bin-width $\Delta\alpha_{CP}$ = $30^\circ$. In each panel the  intersecting points of the black contour  with the one, two and three $\sigma $ contours   depict the respective mixing angles $\phi_\tau$ with SM like Yukawa coupling strength. Shaded interiors of the respective contours are however remain insensitive to the collider.}
\label{fig:FCC_Ep50TeV_Ee150GeV}
\end{figure*}
\section{$CP$-odd Observable}
\label{obsv}
In this section we discuss the observable that are well suited to explore the $CP$ nature of the $h\tau{\bar\tau}$ coupling at the LHeC and FCC-eh. As discussed in Introduction, the major challenge comes from the neutrinos produced from decay of $\tau$-leptons and/ or the additional neutrino in the forward direction produced in association with the Higgs Boson for the charged current. These neutrinos escape detection and their presence can only be deduced from the momentum imbalance which engender measurement of the large missing transverse momenta. 

\par There are various methods exist in the literature   ~\cite{Ellis:1987xu, Elagin:2010aw, Xia:2016jec, Gripaios:2012th, Maruyama:2015fis, Swain:2014dha, Konar:2015hea, Konar:2016wbh} to reconstruct the $\tau$-lepton momentum to construct $CP$ sensitive observables at the LHC. However, in the present analysis at the LHeC/ FCC-eh the presence of additional neutrino in the charged current process makes it extremely challenging to reconstruct the $\tau^\pm$ momenta. Therefore, we do not attempt any reconstruction of $\tau^\pm$ momenta instead utilise the charged and neutral pion's momenta to compute a $CP$ sensitive observable in the charged pion's CM frame   ~\cite{Berge:2008dr}.   

\par To scrutinise the  nature of the $\tau$-lepton Yukawa coupling, we  simulate the $\tau^+-\tau^-$  events corresponding CC and NC processes and allow them to decay  in $\rho^\pm$ mode for LHeC and FCC-eh in \texttt{MadGraph5}. Our method is based  on analysing the acoplanarity angle of the two planes, spanned by $\rho^+$ and $\rho^-$ decay products respectively and defined in $\pi^+-\pi^-$ rest frame known as Zero Momentum Frame (ZMF) as displayed in Fig.~\ref{fig:decayPlane}. Accordingly, on boosting  the momentum of the simulated charged and neutral pions in ZMF of charged pions $\pi^+$ and $\pi^-$,  $CP$ sensitive observable is defined as
\begin{equation}
\alpha_{CP}^\prime \equiv \arccos\left (\hat{p}_{0\perp}^{+} \,\cdot\, \hat{p}_{0\perp}^{-}\right)\,\,\times\,\, {\rm sgn} \left(\hat{p}_{\pi^-}\, \cdot\, \left(\hat{p}_{0\perp}^{+} \times \hat{p}_{0\perp}^{-}\right)\right), \label{alphacpdef}
\end{equation}
where the unit three momentum vectors $\hat p_{\pi^{\pm}}$  specify the directions of the charged pions in ZMF  and $\hat p_{0\perp}^{-}\, \left(\hat p_{0\perp}^{+}\right)$ is the unit   transverse component of the three momentum for neutral pion {\it w.r.t.}  the direction of accompanying charged pion $\pi^-\,\left(\pi^+\right)$  in  ZMF. From equation~\eqref{alphacpdef} it follows that $0^\circ\le \alpha_{CP}^\prime\le 360^\circ$.

\par The  nature of the Yukawa coupling  is hidden in the correlations among the spin vectors $s^-$  and  $s^+$ corresponding to $\tau^-$ and $\tau^+$ in their respective rest frames for which  the partial decay width of $h$ is expressed as
\bea
\Gamma\left(h\to \tau^+\tau^-\right) \propto 1- s^-_{||}\,\, s^+_{||}\,\,\pm \,\,C\,\, s^-_\perp\,\, s^+_\perp, \label{tauspincor}
\eea
where $C$ is complex and unitary. The correlation term for the transverse component in equation~\eqref{tauspincor}  is real and positive (negative) for pure CP even (odd) state but complex for a mixed state. The parity information of $h$ is further encoded in correlations among  the decay products of the $\tau$s  confined in the planes $\perp$ to $\tau^+$ and $\tau^-$ axes. However, the destructive interference among the three  polarised states of the intermediate particle $\rho^\pm$ smear the sensitivity of $\alpha_{CP}^\prime$ in equation \eqref{alphacpdef} due to the modified  correlation term for  transverse component of spin vectors  in equation~\eqref{tauspincor}. Therefore, to unfold the information of a mixed state  we  further impose a  selection cut (filter) and  divide the simulated $\tau$ decay events into two regions, depending on the sign of $Y_{\tau^-\tau^+}$ \cite{Desch:2003rw,Han:2016bvf,CMS:2020rpr}: 
\begin{eqnarray}
Y_{\tau^-\tau^+} =\left(Y_{\tau^-} Y_{\tau^+}\right) = \left[\frac{E_{\pi^-}-E_{\pi^0_{\tau^-}}}{E_{\pi^-}+E_{\pi^0_{\tau^-}}}\right]\times
\left[\frac{E_{\pi^+}-E_{\pi^0_{\tau^+}}}{E_{\pi^+}+E_{\pi^0_{\tau^+}}}\right]\end{eqnarray}
where $E_{\pi^{\pm}}$ and $E_{\pi^0}$ are the energies of charged and neutral pions in the respective $\tau^\pm$ rest frames. Taking into account the impact of sign of $Y_{\tau^-\tau^+}$ we re-define  the observable $\alpha_{CP}^\prime$ as
\begin{equation}
\alpha_{CP}= 
\begin{cases}
\alpha_{CP}^\prime,& \text{for } Y_{\tau^-\tau^+} \geq 0\\
360^{\circ} - \alpha_{CP}^\prime,& \text{for } Y_{\tau^-\tau^+} < 0. \label{eq5}
\end{cases}
\end{equation}
\par We study and analyse the differential distribution of cross-sections {\it w.r.t.} $CP$ sensitive observable $\alpha_{CP}$ for the available kinetic phase space corresponding to varying mixing angle $-{\pi}/{2}\le \phi_\tau\le {\pi}/{2}$.
%$-\frac{\pi}{2}\le \phi_\tau\le \frac{\pi}{2}$. 
It is  observed that  due to adoption of a common decay procedure for $h$ decay to a pair of $\tau^\pm$ and further $\tau^\pm$ decaying in $\rho^\pm$ mode in simulating CC and NC processes, the   shape profile of the normalized differential distributions $({1}/{\sigma})\,({d\sigma}/{d\alpha_{CP}})$ %$\frac{1}{\sigma}\,\frac{d\sigma}{d\alpha_{CP}}$
are found to be similar. The shape profile of normalized distributions for LHeC and FCC-eh operating at two different CM energy are also found to be same as the $\alpha_{CP}$ in ZMF  is designed to be responsive only to the mixing angle $\phi_\tau$. For an illustration, in Fig.~\ref{fig:distLHeC} we display the normalized differential cross-sections  for CC process corresponding to   three set of parameters discussed in section~\ref{sec2:TheorysetUpAndSimulation} for a 150~GeV  $-$80\% polarised electron beam colliding with a 7~TeV unpolarised proton beam at LHeC setup.  The red, green and blue shaded histograms in Fig.~\ref{fig:distLHeC} corresponds to three choices of parameter sets: $(\tilde a =1,\, \tilde b=0)$, $(\tilde a =1/\sqrt{2},\, \tilde b=1/\sqrt{2})$   and $(\tilde a =0,\, \tilde b=1)$  respectively. 
\par The $\alpha_{CP}$  differential distributions are also studied for  dominant irreducible  backgrounds in CC and NC channels,  where $Z/\gamma^\star$ produced through gauge Boson fusion decays    to a pair of $\tau^\pm$ which then further  decay to pions and neutrino/ anti-neutrino in $\rho^\pm$ mode.   In Fig.~\ref{fig:distLHeC} the flat golden yellow histogram depicts the contribution from $Z$ background implying that the constructed $CP$ observable is insensitive to the $\tau^\pm$ decays. 
%%%%%%%%%%%%%%%%%%%%%%%%%%%%%%%%%%%%%%%%%%%%%%%%%%%%%%%%%%%%%%%%%%%%%%%%%%%%%%%%%%%%%%%%%%%%%%%%%%%%%

\section{Results and Discussions}
\label{Results}
In order to estimate the sensitivity of the observable $\alpha_{CP}$ to constrain the $CP$ mixing angle of the $h\tau{\bar\tau}$ coupling we have defined $\chi^2$ as, 
\begin{eqnarray}
\chi^2 \left(\tilde{a},\tilde{b}\right) =\sum_{k = 1}^{n}\left(\frac{ N_k^{\phi_\tau=0}  -  N_k^{\phi_\tau\ne 0}}{\delta  N_k^{\phi_\tau=0}} \right)^2;    \,\,\,\, \left(\tilde a > 0\right),
\label{phicp_sensitivity_chisq}
\end{eqnarray}
where
\begin{eqnarray}
\delta  N_k^{\phi_\tau=0} &= \sqrt{ N_k^{\phi_\tau=0} \left(1 + \delta_{sys}^2 \,\, N_k^{\phi_\tau=0} \right)}.
\end{eqnarray}
Here  $k^{\rm th}$ bin events $\,\,\,\,N_k^{\phi_\tau=0}\,\,\equiv\,\, N_k^{\phi_\tau=0} \left(\tilde{a}=1,\,\,\,\tilde{b}=0\right)$ and $N_k^{\phi_\tau\ne 0}\,\,\equiv\,\, N_k^{\phi_\tau\ne 0}\left(\tilde{a}\ne 0,\,\,\,\tilde{b}\ne 0\right)$ are  the number of  pure $CP$-even state (SM prediction) and $CP$-mixed state events respectively for a given  integrated luminosity $\cal L$. The $\delta_{sys}$  is an approximate  systematic error which includes the luminosity uncertainty. 
\begin{table*}[t]
		\centering
		\resizebox{0.85\linewidth}{!}{
			{\tabulinesep=5pt
				\begin{tabular}{|c|c|c|c|c|c||c|c|c|c|c|c|}
					\cline{1-12}
					\multicolumn{1}{|c|}{\multirow{3}{*}{Process}}&{\multirow{3}{*}{Pol.}}&{\multirow{3}{*}{CL}} & \multicolumn{3}{c||}{LHeC with $L$ = 1~ab$^{-1}$} & \multicolumn{3}{c|}{FCC-eh with $L$ = 1~ab$^{-1}$} & \multicolumn{3}{c|}{FCC-eh with $L$ = 10~ab$^{-1}$}\\
					%\cline{4-12}
					&& & \multicolumn{3}{c||}{$\delta_{sys}$} & \multicolumn{3}{c|}{$\delta_{sys}$} & \multicolumn{3}{c|}{$\delta_{sys}$} \\
					\cline{4-12}
				    && & 0$\%$ & 5$\%$ & 10$\%$ & 0$\%$ & 5$\%$ & 10$\%$ & 0$\%$ & 5$\%$ & 10$\%$ \\
					\cline{1-12}
					\multirow{6}{*}{CC }&\multirow{3}{*}{0\%} &$1\sigma$ & 16.8$^\circ$ & 16.9$^\circ$ & 17.1$^\circ$ & 8.3$^\circ$ & 8.5$^\circ$ & 9.0$^\circ$ & 2.6$^\circ$ & 3.1$^\circ$ & 4.3$^\circ$ \\
					%\cline{3-12}
					& &$2\sigma$ & 32.6$^\circ$ & 32.8$^\circ$ & 33.4$^\circ$ & 17.0$^\circ$ & 17.5$^\circ$ & 18.7$^\circ$ & 5.1$^\circ$ & 6.1$^\circ$ & 8.6$^\circ$
					\\
					%\cline{3-12}
					& &$3\sigma$ & -- & -- & -- & 23.7$^\circ$ & 24.4$^\circ$ & 26.4$^\circ$ & 6.7$^\circ$ & 8.1$^\circ$ & 11.4$^\circ$
					\\
					\cline{2-12}
					&\multirow{3}{*}{$-$80\%} &$1\sigma$ & 11.5$^\circ$ & 11.6$^\circ$ & 12.0$^\circ$ & 6.8$^\circ$ & 7.0$^\circ$ & 7.9$^\circ$ & 2.1$^\circ$ & 2.9$^\circ$ & 4.4$^\circ$ \\
					%\cline{3-12}
					& &$2\sigma$ & 24.5$^\circ$ & 24.9$^\circ$ & 26.0$^\circ$ & 13.7$^\circ$ & 14.3$^\circ$ & 16.0$^\circ$ & 4.2$^\circ$ & 5.6$^\circ$ & 8.8$^\circ$ \\
					%\cline{3-12}
					& &$3\sigma$ & 36.6$^\circ$ & 37.4$^\circ$ & 39.9$^\circ$ & 18.6$^\circ$ & 19.5$^\circ$ & 22.1$^\circ$ & 5.5$^\circ$ & 7.5$^\circ$ & 11.7$^\circ$ \\
					\cline{1-12}
					
					\multirow{6}{*}{NC} &\multirow{3}{*}{0\% }& $1\sigma$& -- & -- & -- & -- & -- & -- & 5.8$^\circ$ & 6.1$^\circ$ & 7.0$^\circ$ \\
					%\cline{3-12}
					&& $2\sigma$ & -- & -- & -- & -- & -- & -- & 11.6$^\circ$ & 12.3$^\circ$ & 14.1$^\circ$ \\
					%\cline{3-12}
					&& $3\sigma$ & -- & -- & -- & -- & -- & -- & 15.7$^\circ$ & 16.7$^\circ$ & 19.3$^\circ$ \\
					\cline{2-12}
					&\multirow{3}{*}{$-$80\% } & $1\sigma$ & -- & -- & -- & -- & -- & -- & 5.6$^\circ$ & 5.9$^\circ$ & 6.8$^\circ$\\
					%\cline{3-12}
					&& $2\sigma$ & -- & -- & -- & -- & -- & -- & 11.0$^\circ$ & 11.6$^\circ$ & 13.4$^\circ$\\
					%\cline{3-12}
					&& $3\sigma$ & -- & -- & -- & -- & -- & -- & 14.6$^\circ$ & 15.5$^\circ$ & 17.9$^\circ$\\
					\cline{1-12}
				\end{tabular}}
			}
			
\caption{{\small Figure of merit displaying the lower limits on the  mixing angle $\phi_\tau $ upto which the proposed LHeC and FCC-eh colliders can probe with   68\%, 95\% and 99.7\% C.L. based on the  $\chi^2$ analysis for CC and NC processes. The sensitivities are computed using  differential   bin-width $\Delta\alpha_{CP}$ = $30^\circ$ corresponding to three choices of the systematic errors for a given integrated  luminosity. } }
			\label{tab:sensitivitytable}   
		\end{table*}
\par Assuming that the deviations in the number of events from the SM predicted pure $CP$-even state are due to  variation of either the strength of the coupling or the mixing angle $\phi_\tau\ne 0$, we perform the $\chi^2$  analysis with  the  histograms drawn from the parton level one  dimensional differential cross-sections {\it w.r.t.} $\alpha_{CP}$  by varying $\tilde a$ and $\tilde b$. Using the differential bin-width $\Delta\alpha_{CP}=30^\circ$, the $\chi^2$ is computed for one degree of freedom in the two-dimensional plane spanned by $\tilde a- \tilde b$. 

\par The one $\sigma$ (in red) and two $\sigma$ (in blue) exclusion contours in $\tilde a-\tilde b$ plane are depicted in Fig.~\ref{fig:CC_Ee150} for the charge current process at LHeC with 150~GeV  electron beam and 7~TeV  proton beam with zero systematic error and an integrated luminosity of $\cal L$ = 1 ab$^{-1}$. The left and right panels in Fig.~\ref{fig:CC_Ee150} correspond to the unpolarised and $-$ 80\% polarised electron beams respectively. The unshaded exterior regions corresponding to respective contours  can be probed by the proposed LHeC collider. Restricting the magnitude of the coupling to be SM like, we draw a contour  $\sqrt{\tilde a^2 + \tilde b^2}$ = 1 depicted in black which intersect the three $\chi^2$ contours. The two intersecting coordinates define the  respective lower limits of the mixing angle $\tan^{-1}\left(\tilde b/\,\tilde a\right)$ that can be measured in the proposed collider at 68\% and 95\%  C.L. respectively. These limits on the mixing angles  for  the unpolarised and $-$80\% polarised electron beams are given in Table~\ref{tab:sensitivitytable}. The $\chi^2$ analysis 
are also performed with non-zero systematic errors for  $\delta_{sys}$ of the order of  5$\%$ and 10$\%$ respectively with the same differential distributions and the corresponding relaxed  lower limits of the mixing angle  are given in Table~\ref{tab:sensitivitytable} at 1 $\sigma$, 2 $\sigma$ and 3 $\sigma$ C.L. respectively.

\par The $\chi^2$ analysis for $\alpha_{CP}$ distributions from NC process are also performed with unpolarised and polarised beams respectively  but the corresponding sensitivity limits on the mixing angle are found to be rather weak due to comparatively small cross-sections as shown in the Table~\ref{tab:xsec}. 
\begin{figure}[t]
\includegraphics[width=0.5\textwidth]{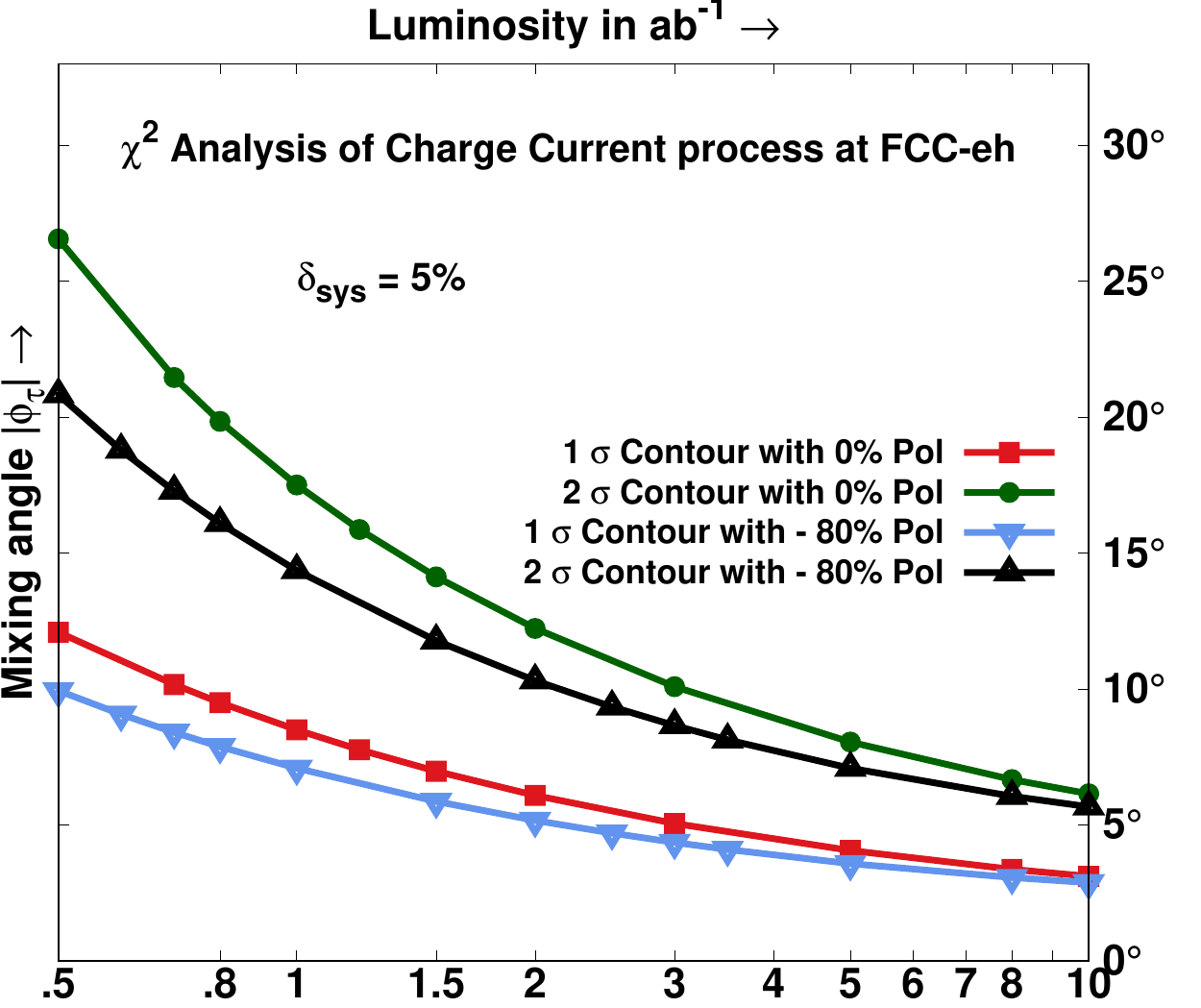}
\caption{  {\small $\chi^2$ contours are drawn in luminosity - lower limit on the Mixing angle plane for unpolarised ($-$80\% polarised) corresponding to the contribution from charged current process in the FCC$-$eh setup with $\delta_{sys} = 5\%$.}}
	\label{fig:sensitivity_vs_lumi}
\end{figure}
\par Similarly in Fig.~\ref{fig:FCC_Ep50TeV_Ee150GeV} we have displayed the one $\sigma$ (in red), two $\sigma$ (in green) and three $\sigma$ (in blue) exclusion contours in  $\tilde a-\tilde b$ plane for the FCC-eh set up with electron and proton beam energy of 150~GeV and 50~TeV, respectively.  The first two figures on the upper panel  are drawn from the contributions of CC process where the  left panel  and right panel correspond to the unpolarised and $-$80\% polarised electron beams respectively with zero systematic error and an integrated luminosity $\cal L$ = 10 ab$^{-1}$. The lower panel of two figures depict the contours drawn from the differential distributions of the sub-dominant NC process  corresponding to the unpolarised and $-$80\% polarised electron beams, respectively with zero systematic error and an enhanced integrated luminosity $\cal L$ = 10 ab$^{-1}$. In all the four panels we draw the black colour contour for $\tilde a^2 +\tilde b^2$ = 1 to extract the  lower limits of mixing angles that can be measured in the proposed FCC-eh collider at 68\%, 95\% and 99.7\% C.L. respectively.  These lower limits on mixing angles at the 1 $\sigma$, 2 $\sigma$ and 3 $\sigma$ C.L. are given in Table~\ref{tab:sensitivitytable} for two choices of integrated luminosities $\cal L$ = 1 and 10~ab$^{-1}$ and three representative values  of  systematic errors (0\%, 5\% and 10\%) corresponding to each luminosity choice.\footnote{Due to the small cross section and computational limitations in case of Fig.~\ref{fig:CC_Ee150}, we restrain ourselves to provide the 3\,$\sigma$ contour. For similar reasons, we have not provided the 3\,$\sigma$ limits in Table~\ref{tab:sensitivitytable} for unpolarised charged current process.}
 
\par Finally we analyse $\chi^2$ as a function of the  luminosity and the lower limit on the mixing angle upto which the proposed FCC-eh collider  can be sensitive at a chose value of  C.L.   Fig.~\ref{fig:sensitivity_vs_lumi} display the 68\% (red and blue) and 95 \% (green and black)  C.L. contours  in luminosity - lower limit of the mixing angle plane   with $\delta_{sys}$ 5\%, for the dominant charge current contribution  at FCC-eh set up corresponding to unpolarised  and $-$80\% polarised  150~GeV electron beams, respectively.

\subsection{Contamination due to $Z/\gamma^*$ background}
\label{BGZGammaContrib}

\begin{figure*}[t]
\includegraphics[width=0.5\textwidth]{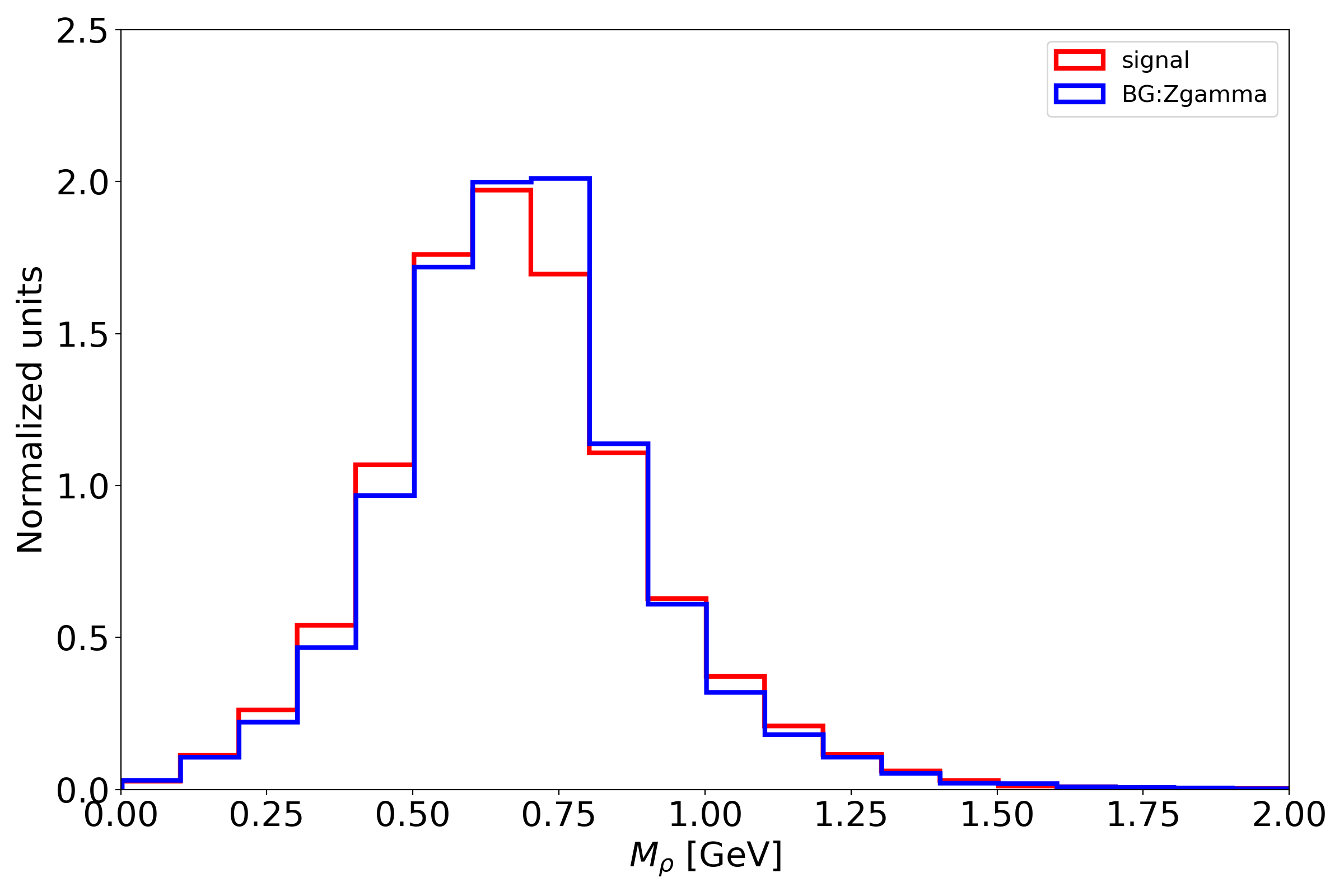}
\includegraphics[width=0.5\textwidth]{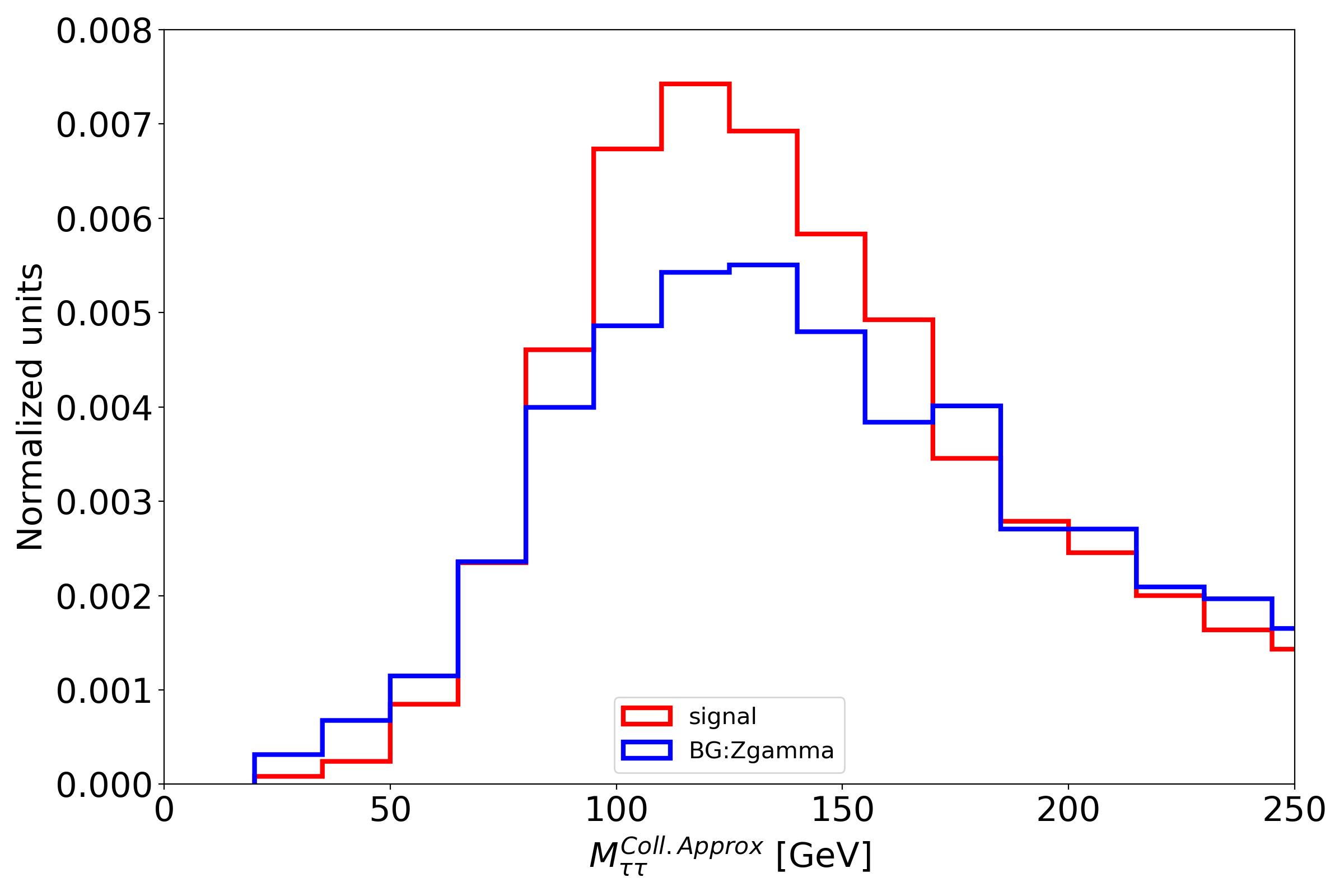}
\caption{  {\small Left panel displays the distribution of the rho mass computed from the charged pions and the neutral pions. In the right panel we show the reconstructed $\tau$ pair invariant mass using the collinear approximation. Clearly the collinear approximation works well in getting the Higgs boson mass at the correct value.}}
	\label{fig:RhoMassMtautauInvMass}
\end{figure*}
We've discussed about the results at the parton level so far, and computed the sensitivity of the $CP$ sensitive observable, $\alpha_{CP}$, to constrain the $CP$ admixture of the spin-0 mediator decaying to pair of $\tau$ leptons. However, if we add the potential backgrounds with realistic effects owing to showering, hadronization, and detector architecture, these results are likely to be altered. We'll go through each of them briefly in this sub-section.

\par In the $CP$ study of scalar boson at $e^-p$ collider, we strategize to minimise contamination from the dominant background ($\sim$ six times larger than signal) where $\tau$-leptons are produced from $Z$-boson and/or $\gamma^\star$ decay in addition to the aforementioned background of the SM Higgs boson $0^+$ state decaying to di-tau. Background contamination from the other potential sources contributing to formation of prompt soft pions is however, negligible for ${p_T}_{\pi^{\pm,0}}\ge$ 20~GeV.\footnote{In addition, this cut is also very important in controlling the soft charged tracks inside the $\tau$-jet radius which may affect the shape of the $\alpha_{CP}$ observable.}

\par In order to compute the efficiency with which we can discriminate and minimize the $Z^0/\,\gamma^\star$ background from the decay products of spin-0 state, we employ a deep neural network (DNN) learning algorithm consisting  of one input layer with 20 nodes and one output layer with activation function \texttt{sigmoid}. In addition, we have three hidden layers with nodes corresponding to each layers are 200, 200 and 20 respectively. The activation function for the input and the hidden layers are chosen to be \texttt{relu}. We then optimized the DNN where  the learning rate, $\beta_1$, $\beta_2$ and $\epsilon$ were set to  their default values as 0.0005, 0.9, 0.999, $1e^{-08}$ respectively. We have chosen \texttt{Adam} as the optimizer and \texttt{binary crossentropy} as the loss function. We split the data set for training and testing as 70\% and 30\% respectively with 30 epochs and batch size of 50. 

\par The network's input kinematic variables  are combination of both the low level and the high level observables. 
The low-level inputs are the four momenta of the final state charged and  neutral pions produced from the decays of $\tau$-leptons, whereas the high-level input observables are $\Delta \phi$, $\Delta R=\sqrt{(\Delta \eta)^2 + (\Delta \phi)^2}$ and invariant mass of reconstructed $\rho^\pm$ from the observed four pions, reconstructed $\tau$ pairs and missing transverse momenta. Since both $m_\tau/ m_Z$ and $m_\tau/ m_{h^0}$ are $\ll$ 1, we assume that $\tau$'s are  highly boosted and therefore, we have adopted the collinear approximation to reconstruct the invisible neutrino momenta and then reconstruct the $\tau$-pair invariant mass. Now, among all the input variables to the network, we removed the ones that are highly correlated with others. Here, we utilise \texttt{Pearson} correlation to compute the correlation among the input variables.  Following that, the network is fed with a total of 23 kinematic observables. With this, we can separate the signal from the background with an accuracy of 83\%. All of these findings are for the parton level sample, and the details may be seen in the Appendix below.

\par We find that the DNN  enhances the sensitivity of  the $CP$ phase  determination in the $\chi^2$ analysis and further lowers the one sigma limit to 16.0$^\circ$ for polarized LHeC with luminosity of  1~ab$^{-1}$. The results for other cases are also similarly affected. In the following we will discuss the effects of the detector simulation on these sensitivities.

\subsection{Detector Simulation}
\label{DetectorSimulation}

%{fig:RhoMassMtautauInvMass}
\begin{figure}[t]
\includegraphics[width=0.45\textwidth]{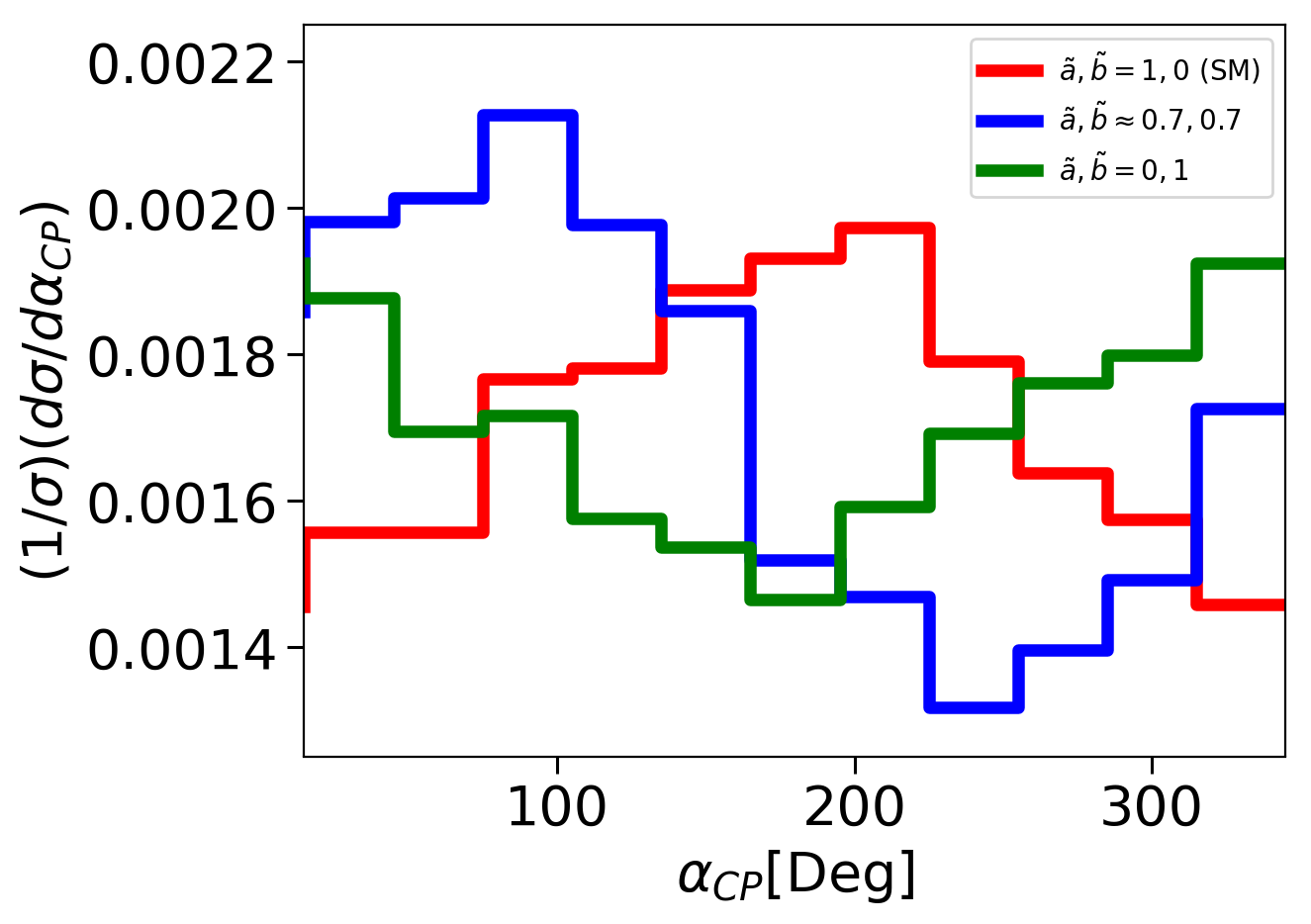}
\caption{  {\small Here we present the distribution of the $\alpha_{CP}$ after the detector simulation. The red, blue and green distributions represent the $CP$-phase zero, $\pi/4$ and $\pi/2$ respectively. Evidently the $\alpha_{CP}$ observable is very important in constraining the $CP$-phase.}}
	\label{fig:alphaCPAfterDelphes}
\end{figure}

We have done showering and hadronization using \texttt{Pythia8} \cite{Sjostrand:2006za}. Since \texttt{Pythia8} is still new to handle the $e^- p$ collider environment we have adjusted few settings in a standalone \texttt{Pythia8} code\footnote{We had a private communication with the \texttt{Pythia8} authors and following the suggestion we have switched off the QED radiation from the lepton and also switched off the the input matching of the \texttt{LesHouches} input obtained from \texttt{Madgraph5}.} to enable the showering and hadronization. The detector simulation is performed using \texttt{Delphes3}~\cite{deFavereau:2013fsa}. As the LHeC and FCC-eh are asymmetric colliders the $\eta$ ranges of the final state particles are very different from a symmetric collider like LHC, so we have modified the detector card accordingly to take care of that issue~\cite{Kumar:2015kca}. Additionally, we also have modified the efficiencies and isolation criteria of different objects as per the technical design report of these colliders~\cite{AbelleiraFernandez:2012cc}. The jet construction are done using \texttt{Fastjet}~\cite{Cacciari:2011ma} which utilize anti$-k_T$ algorithm with radius R = 0.5 and $p_T > 20$~GeV.

The events are selected with at least two $\tau$ jets with $p_T > 20$~GeV and we have not put any cuts on $\eta$ because of the asymmetric nature of the collider. Once we have get the $\tau$ jets we get the $\tau$ constituents from which tracks gives the charged pions and the tower corresponds to the neutral pions. Since in most of the cases there are many charged tracks and also many entries in the tower, so one should make sure that the right combination of the charged track and tower should be selected such that the invariant mass is close to the $\rho$ meson mass (770~MeV). In Fig.~\ref{fig:RhoMassMtautauInvMass} left panel we display the reconstructed $\rho$ meson mass obtained from the charged and neutral pions four momenta that we get from the track and tower. %Evidently, most of the events are having the right $\rho$ mass and hence we hav right rho mass. 

With the correct combination of charged and neutral pions, we can reconstruct the neutrinos to obtain the $\tau$ lepton pair invariant mass. For the hadronic decays of the $\tau$'s, two neutrinos are present in each event, which are boosted and are collinear with respect to their parents. The collinear approximation, however, assumes that there is no other source of the missing transverse energy other than neutrinos. But in the charged current mechanism we studied here, the forward neutrino provides an additional source of missing transverse energy, causing the invariant mass to have a large tail.

\par In the right panel of Fig.~\ref{fig:RhoMassMtautauInvMass}  we display the $\tau$ pair invariant mass which is peaking at the Higgs bososn mass. We observe that the forward neutrino contribution to the missing energy from the spin-1 background is more prominent than the signal.

\par To maximise the signal to background ratio, we use the same deep learning network structure and train with the same observable as in the parton-level analysis, resulting in a 73\% accuracy. For both testing and training samples, the area under the curve (AUC) of the receiver operating characteristic (ROC) curve is 80\%. The distributions of the $\alpha_{\rm CP}$ are shown in Fig.~\ref{fig:alphaCPAfterDelphes} with red, blue, and green colour histograms, which correspond to the $CP$ phases zero, $\pi$/4, and $\pi$/2, respectively. The distributions for the FCC-eh polarised collider are shown here.

Using the $\chi^2$ analysis that we described before, we get 23.32$^\circ$ at 1$\sigma$ for 3~ab$^{-1}$ luminosity and  46.0$^\circ$ at 2$\sigma$ for this $CP$ sensitive observable. We can further improve and constrain the $CP$-phase to  11.4$^\circ$ at 1$\sigma$ and 24.0$^\circ$ at 2$\sigma$ for polarized electron beam at FCC-eh.

\section{Summary and Conclusion}
\label{summaryAndConclusion}

\par In this article we explore the $CP$ mixing probability of $h$ through its decay to $\tau^+\tau^-$ in the future $e^-p$ collider, where $h$ is singly produced in the charged and neutral current modes through $W^+W^-$ and $ZZ$-fusion, respectively. 
To explore the $CP$ nature through the $h\tau\bar\tau$ coupling, we consider the $\tau^{\pm} \to \pi^{\pm} + \pi^{0} + \bar\nu_\tau/\nu_\tau$ decay modes as this channel has largest branching fraction of about 25\%.   
We have deployed an interesting observable, $\alpha_{CP}$, which does not require reconstruction of $\tau^\pm$ to scrutinise the $CP$ sensitivity of $h\tau\bar\tau$ coupling. With this observable we employed $\chi^2$ between the SM expectation and new physics to estimate the sensitivity. 

\par The sensitivity that is obtained from this analysis with charged current at the LHeC with $\cal{L} =$ 1~ab$^{-1}$ is 17$^\circ$ (12$^\circ$) for unpolarised (polarised) electron beam at 68$\%$ C.L. when uncertainty is 10$\%$. The sensitivity at 95$\%$ C.L. is 33$^\circ$ (26$^\circ$) for unpolarised (polarised) electron beam. Expectedly, the best limit is obtained from the FCC-eh with charged current process at $\cal{L} =$ 10~ab$^{-1}$, 4$^\circ$ at 68$\%$ C.L. while 9$^\circ$ at 95$\%$ C.L. with 10$\%$ uncertainty. Similarly, the sensitivity for FCC-eh at 1~ab$^{-1}$ is 9.0$^\circ$ (8.0$^\circ$) and 18.7$^\circ$ (16.0$^\circ$) at 68$\%$ and 95$\%$ C.L. respectively for unpolarised (polarised) electron beam respectively. The limit for the neutral current is very weak for LHeC setup and even for FCC-eh and hence the limits are not shown here. Hence, the sensitivity for neutral current is shown for FCC-eh with $\cal{L} =$ 10~ab$^{-1}$. This limit is 7.0$^\circ$  and 14.0$^\circ$ at 68$\%$ and 95$\%$ C.L. respectively. Here also we have considered 10$\%$ uncertainty in the computation. Hence, it is evident from this analysis that the $CP$ phase of $h\tau\bar\tau$ coupling can be measured quite efficiently in a futuristic $e^-p$ collider and the sensitivity might be comparative, if not better, compared to the limit obtained from LHC. Although at the LHC the production cross section is higher than what we can expect at a $e^-p$ collider, in the respect of backgrounds $e^-p$ machine has low backgrounds making the limits obtained from them are comparable with LHC. 

Considering the effects of background(s) and smearing at the detector-level the sensitivities of $CP$-phase observable is affected and studied. With the $\chi^2$ analysis using the $CP$-phase observable, the $CP$ admixture of the $h\tau\bar{\tau}$ coupling is constrained to 23.32$^\circ$ (46.0$^\circ$) at 1 (2) $\sigma$ for 3~ab$^{-1}$ luminosity in the case of FCC-eh where $E_e = 150$~GeV and $E_p = 50$~TeV. However, at ${\cal L} = 10$~ab$^{-1}$ it is 11.4$^\circ$ (24.0$^\circ$) at 1 (2) $\sigma$.

\par It is important to mention that in this study $h\tau{\bar\tau}$ is taken to be the only $CP$-violating coupling, all other couplings used in this work are the SM couplings.Though the $CP$ nature of $hWW/hZZ$ couplings has been studied in literature~\cite{Biswal:2012mp,Kumar:2015kca} and one can also perform a global analysis considering the $CP$ structure at production as well as at the decay vertex. The $CP$ nature of $h$ can also be probed through its production in association with top-quarks   ~\cite{Coleppa:2017rgb} and there one can simultaneously account the $CP$ admixture at $h$-decay vertex via $h\to\tau^+\tau^-$. This channel also open a space to probe simultaneous measurement of $hWW$, $Wtb$, $ht\bar t$ and $h\tau\bar\tau$ couplings and we keep these possibilities for future studies.

\section*{Acknowledgements}
The work of SD, AG and AKS was partially supported by the SERB, Govt. of India under  CRG/2018/004889. We thank Satyaki Bhattacharya, Kai Ma and Ilkka Helenius for useful discussion.

%\section*{References}
%\appendix
%\label{append}
\section*{Appendix}
\begin{figure*}[t]
\centering
\includegraphics[width=0.49\textwidth]{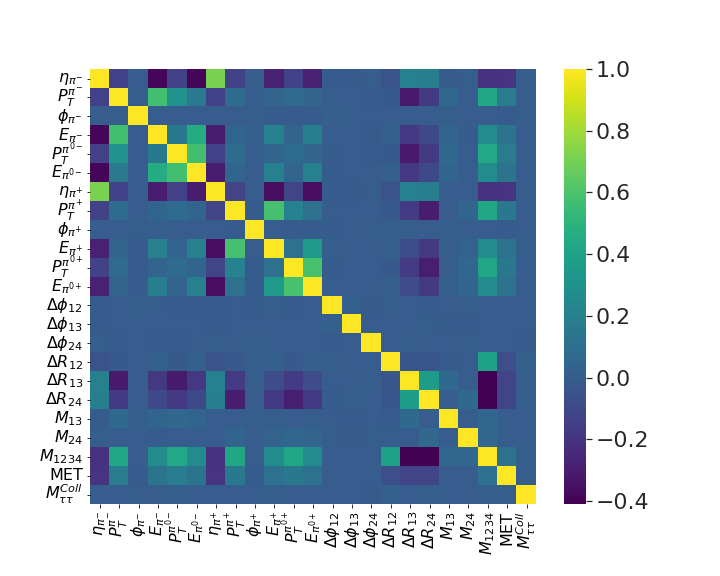}
\includegraphics[width=0.49\textwidth]{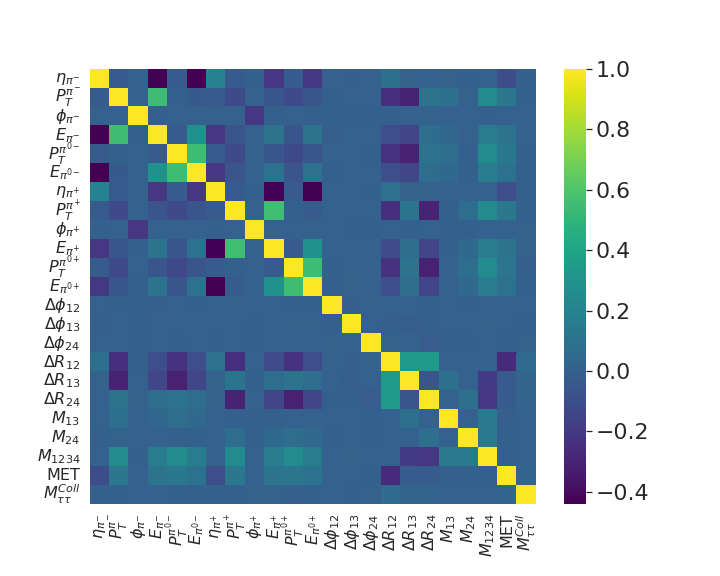}
\caption{  {\small Correlation among the kinematic observables (both low level as well as high level), which are used as input to the deep learning algorithm is displayed here. The left and right panel corresponds to the background and signal events, respectively.}}
	\label{fig:ObsCorrelationBGSIG}
\end{figure*}
\begin{figure*}[t]
\centering
\includegraphics[width=0.35\textwidth]{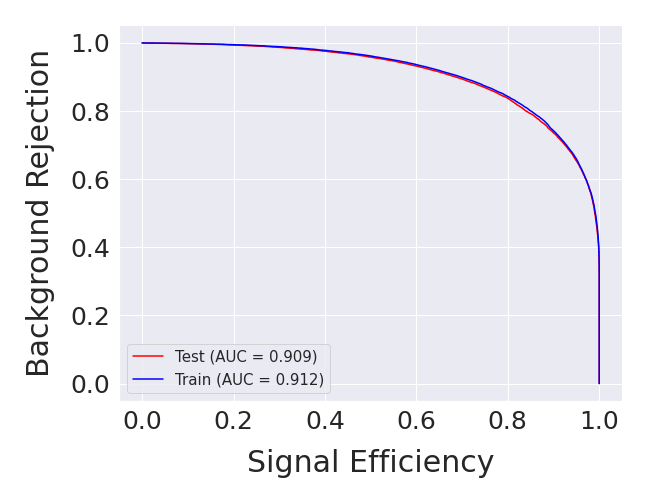}
\includegraphics[width=0.53\textwidth]{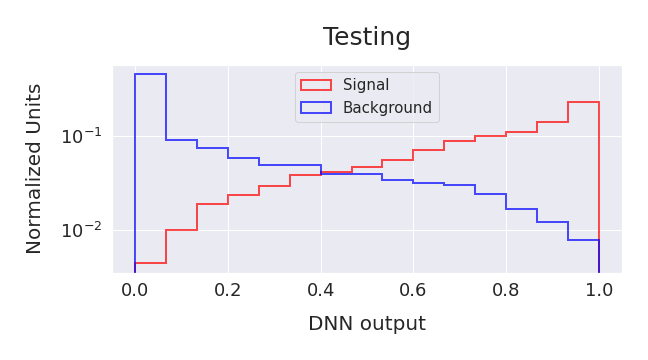}
\caption{  {\small Left panel: the ROC curve. Right panel: the DNN output which is used to discriminate the signal and background events.}}
	\label{fig:ROCandDeepOutputpartonlevel}
\end{figure*}
Here we discuss about the deep learning algorithm that we employed in this analysis. The basic kinematic observables like transverse momenta ($p_T$), pseudo rapidity ($\eta$), azimuthal angle ($\phi$) and energy ($E$) of each charged and neutral pions are referred as low level variables. The high level variables are the ones constructed from the low level variables such as $\Delta \phi_{ij}$, $\Delta R_{ij}$, invariant mass ($M_{ij}$) etc. Here $i,j = \pi^-, \pi^+,\pi^{0-}, \pi^{0+}$ with $0\pm$ in the superscript refers to the neutral pions produced from $\tau^{\pm}$.

In Fig.~\ref{fig:ObsCorrelationBGSIG} we show the correlation among the kinematic observables, input to the deep learning algorithm. The left (right) panel corresponds to background (signal) events. There few other kinematic variables which can be there but we have removed them from the list because of the high correlation with other existing variables. Both for the signal and the background events the correlation among different variables are more or less similar.

We also show here the ROC curve and the deep learning algorithm output to discriminate the signal and background in Fig.~\ref{fig:ROCandDeepOutputpartonlevel}. This is done at the parton level with $CP$ phase 0 of Higgs to $\tau$ pair sample as the signal. We have done the same exercise for the other $CP$ phase values like $\pi/4$ and $\pi/2$ and found that the result remains the same. This is essentially because the kinematic observables are more or less insensitive to the $CP$ phase of the Higgs to $\tau$ pair samples. The right panel describes the deep learning output for the signal and background events on the testing sample with which the algorithm has not been trained. It is also evident from this that the deep learning algorithm learned the differences between the signal and background kinematics and successfully separated them. At the detector level we also observed the similar separation of signal and background events from the DNN algorithm used in this work.

\end{document}